\documentclass[manuscript,screen,nonacm]{acmart}

\AtBeginDocument{%
  \providecommand\BibTeX{{%
    \normalfont B\kern-0.5em{\scshape i\kern-0.25em b}\kern-0.8em\TeX}}}

\setcopyright{rightsretained}
\copyrightyear{2022}
\acmYear{2022}

\usepackage{xr-hyper}
\usepackage{todonotes}
\usepackage{hyperref}
\usepackage[inline]{enumitem}

\usepackage{graphicx}

\usepackage{multirow}
\usepackage{xcolor,soul,colortbl}
\usepackage{longtable}

\definecolor{Gray}{gray}{0.85}

\makeatletter
\newcommand*{\addFileDependency}[1]{
  \typeout{(#1)}
  \@addtofilelist{#1}
  \IfFileExists{#1}{}{\typeout{No file #1.}}
}
\makeatother

\newcommand*{\myexternaldocument}[1]{%
    \externaldocument{#1}%
    \addFileDependency{#1.tex}%
    \addFileDependency{#1.aux}%
}

\myexternaldocument{2_RelatedWork}

\begin{document}

\title{WoW -- A System for Self-Service Collaborative Design Workshops}






\author{\textbf{Ilyasse Belkacem}}
\affiliation{%
  \institution{Luxembourg Institute of Science and Technology}
  \streetaddress{5 Avenue des Hauts-Fourneaux}
  \city{4362 Esch-sur-Alzette}
  \country{Luxembourg}}
\email{ilyasse.belkacem@list.lu}

\author{\textbf{Vasile Ciorna}}
\affiliation{
  \institution{Goodyear Innovation Center Luxembourg}
  \streetaddress{Avenue Gordon Smith}
  \city{L-7750 Colmar-Berg}
 \country{Luxembourg}}
\email{vasile_ciorna@goodyear.com}

\author{\textbf{Frank Petry}}
\affiliation{%
  \institution{Goodyear Innovation Center Luxembourg}
  \streetaddress{Avenue Gordon Smith}
 \city{L-7750 Colmar-Berg}
  \country{Luxembourg}}
\email{frank.petry@goodyear.com}

\author{\textbf{Mohammad Ghoniem}}
\affiliation{%
  \institution{Luxembourg Institute of Science and Technology}
  \streetaddress{5 Avenue des Hauts-Fourneaux}
  \city{4362 Esch-sur-Alzette}
  \country{Luxembourg}}
\email{mohammad.ghoniem@list.lu}

\renewcommand{\shortauthors}{I. Belkacem et al.}

\begin{abstract}
  In many working environments, users have to solve complex problems relying on large and multi-source data.
  Such problems require several experts to collaborate on solving them, or a single analyst to reconcile multiple complementary standpoints.
  Previous research has shown that wall-sized displays supports different collaboration styles, based most often on abstract tasks as proxies of real work.
  We present the design and implementation of WoW, short for ``Workspace on Wall'', a multi-user Web-based portal for collaborative meetings and workshops in multi-surface environments.
  We report on a two-year effort spanning context inquiry studies, system design iterations, development, and real testing rounds targeting design engineers in the tire industry.
  The pneumatic tires found on the market result from a highly collaborative and iterative development process that reconciles conflicting constraints through a series of product design workshops.
  WoW was found to be a flexible solution to build multi-view set-ups in a self-service manner and an effective means to access more content at once.
  Our users also felt more engaged in their collaborative problem-solving work using WoW than in conventional meeting rooms.
\end{abstract}



\begin{teaserfigure}
  \includegraphics[width=\textwidth]{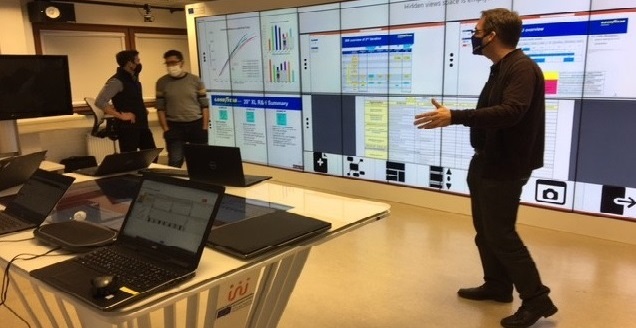}
  \caption{Tire design meeting. The manager is animating the discussion. Multiple documents are shared on the wall display from coworkers' personal devices.  A side discussion is taking place at the left side. }
  \Description{Enjoying the baseball game from the third-base
    seats. Ichiro Suzuki preparing to bat.}
  \label{fig:teaser}
\end{teaserfigure}

\maketitle


\section{Introduction}

Modern companies strive to use the latest technologies and set industry standards amidst high competition. 
Customers are more stringent regarding product performance. 
Although companies collect large amounts of data from their business and manufacturing processes, 
it is hard to leverage this data efficiently to achieve the best products and services.
Despite a trend to acquire `\emph{T-shaped employees}'~\cite{conley2017acquisition}, i.e. with a deep expertise in one field and a broad knowledge in many others, 
it is still hard for individuals to master multiple fields, hence the need for collaboration.
A common office setup comprises a single workstation with two monitors. 
This setup increases user performance and satisfaction compared to single monitor setups~\cite{owens2012examination}.
For collaborative work, the most common setup is a laptop and a projector in a meeting room.
This allows people to be colocated and to freely discuss the projected material. 
But these meetings are often prepared upfront by an animator who leads the discussions.
Other participants will mostly opine about the material on display.
They can share content only if they interrupt the ongoing show.

Large high-resolution displays (LHRDs), often referred to as \textit{Wall-Sized Displays} (WSDs) or \textit{Powerwalls}~\cite{rooney2013powerwall,lischke2015interaction} have recently become more affordable.
Often set up as a grid of high-resolution screens or projectors~\cite{leigh2012scalable}, they 
support user interaction up-close or from a distance~\cite{beaudouin2012multisurface,jakobsen2014up,kister2015bodylenses}.    
Several studies have analyzed the benefits and challenges related to WSDs. 
The benefits include: an increased screen real-estate to display massive data sets as one high-resolution view ~\cite{ruddle2016design} or as multiple coordinated views~\cite{langner2018multiple}; a higher resolution allowing reading up close and getting an overview from a distance~\cite{ball2007move,jansen2019effects}; a larger space in front of the display to fit multiple coworkers and allowing for personal and shared spaces~\cite{prouzeau2016evaluating}; and new interaction possibilities beyond traditional UI paradigms and techniques~\cite{jakobsen2013information,badam2016supporting,liu2017coreach}.  
WSDs were used for data visualization and sensemaking~\cite{kister2017grasp}, workshops and meetings~\cite{thomas2017echo}, command and control~\cite{ion2013canyon}, education~\cite{chattopadhyay2018shared,clayphan2016wild}, 
gaming~\cite{von2016miners,mayer2018pac} and public exhibitions~\cite{peltonen2008s}.
They were also used for
the visualization of gigapixel images in healthcare and space~\cite{ruddle2016design,pietriga2016exploratory}, traffic and infrastructure management~\cite{prouzeau2016towards}, emergency response~\cite{chokshi2014eplan,chan2016envisioning,onorati2015walltweet}, scheduling conferences~\cite{doshi2017stickyschedule}, business process modeling~\cite{nolte2016collaborative}, and product design~\cite{khan2009toward}.

In this paper we introduce WoW, short for `Workspace on Wall', a framework that supports
colocated collaborative work. 
The proposed solution is based on a thorough inquiry, spanning several months, into the potential offered by new multi-surface environments in processes and activities surrounding the design of pneumatic tires.
Designing an efficient tool to support collaborative work in such a context requires a thorough analysis of workflows involving multiple users working together on solving complex problems.

This paper makes three main contributions:
\begin{enumerate*}[label={\arabic*)}]
	\item the elicitation of the needs and workflow of technical tire design workshops;
	\item WoW, a prototype software for collaborative work on large visualization walls; 
	\item A usability study with an industrial work group focusing on a real use case.
\end{enumerate*}
In the rest of this paper,
Section~\ref{sec:rel} discusses the related work;
Section ~\ref{sec:prem} recaps a five-month context inquiry research; 
Section ~\ref{sec:sysdesc} motivates our work; 
Section ~\ref{sec:study} describes the WoW system; 
Section ~\ref{sec:findings} summarizes the usability study with two applied engineering teams from the Goodyear Innovation Center, Luxembourg; 
in Section ~\ref{sec:disc} we discuss our findings. 
Section ~\ref{sec:concl} discusses the limitations and outlook of this work.


\section{Related work}
\label{sec:rel}

\subsection{Collaborative problem solving} \label{sec:CPS}
As noted by Hutchins~\cite{hutchins1995cognition}, one can hardly think of a product which does not result from collaboration.
Solving complex engineering problems also relies on complementary skills and expertise. 
Collaborative problem solving (CPS) is divided into four categories of activities and skills:
\begin{enumerate*}[label={\arabic*)}]
	\item Teamwork;
	\item Communication;
	\item Leadership;
	\item Problem solving~\cite{oliveri2017literature}.
\end{enumerate*}


\textbf{Teamwork} was widely studied in psychology~\cite{cohen1991teamwork}.
Most models of teamwork include dimensions such as adaptability, coordination and cooperation, energizing task efforts and conflict resolution~\cite{driskell2018foundations}.
Recent research emphasizes the poor coverage of
the temporal dynamics of teamwork~\cite{marques2019there,wiese2019understanding,shuffler2020evolution} and of creativity and innovation processes~\cite{benishek2019teams}. 

\textbf{Communication} with its two constituents, active listening and exchanging information~\cite{oliveri2017literature}, is key to unleash synergies making groups perform better than their best individual member~\cite{stasser2020collective}. 
This requires that users share their unique and individual knowledge effectively for solving a problem deemed unsolvable using only participants' common knowledge~\cite{stasser1985pooling,lu2012twenty}.
Also group failures are often communication failures~\cite{stasser1985pooling}.

\textbf{Leadership} as a supportive behavior has direct and indirect consequences on problem-solving~\cite{carmeli2013leadership}. 
New approaches such as empowering leadership~\cite{hassan2013ethical} and authentic leadership~\cite{gardner2005veritable} are found to be the most associated to positive outcomes in innovation and creativity
~\cite{hughes2018leadership}. 
Central to empowering leadership is the concept of sharing information and seeking input while having the authority to make decisions.
One can imagine teams working collaboratively and disposing of enough information may make autonomous decisions, thus being more rapid and efficient. 

\textbf{Problem solving} is a distinguished capability of any team\cite{holyoak1990problem}.
It can be divided into two steps
\begin{enumerate*}[label={\arabic*)}]
	\item constructing the representation of the problem, and 
	\item generating the solution~\cite{novick2005problem}.
\end{enumerate*}
Solution generation is ascribed to the occurrence of insights, which in turn is still a mystery\cite{davidson2003insights,nguyen2009framework}.

\subsection{Collaboration around LHRDs} \label{ssec:colabLHRD}

Mateescu et al.~\cite{mateescu2021collaboration} structure the effects of large displays on \emph{collaborative processes} 
along six axes: workplace awareness, verbal and gestural communication, participation, coordination flow, artifact interaction and level of reasoning. 

Workspace awareness defined as “the up-to-the-moment understanding of another person’s interaction with the shared workspace”~\cite{gutwin2004importance} is a crucial component of team cognition. 
The increased screen real-estate of LHRDs leads to a positive impact on mutual awareness between users while being able to have a private working space~\cite{clayphan2016wild,onorati2015walltweet}. 
LHRDs raise different design opportunities and solutions to maintain mutual awareness while interacting with the content displayed on the screen. 
For instance, Prouzeau et al.~\cite{prouzeau2018awareness} display information about other people's actions to support collaboration in a crisis management scenario.
Besides common verbal communication, users frequently communicate through gestures, body postures, facial expressions and eye contact~\cite{langner2018multiple,Zadow2016miners}. 

In a conventional meeting room with a laptop and projector, the meeting is usually led by the person controlling the laptop while other participants observe.
Due to the lack of participation equity, important information may not be communicated, thus increasing the chance of poor decisions~\cite{dimicco2004influencing}. 
In shared working environments, participation equity can be enabled, encouraged and affected by the availability of a shared display space and by the input configuration~\cite{rogers2009equal}. 
LHRDs can promote equal participation in groups without necessarily having to be encouraged, or constrained through enforced turn taking~\cite{jakobsen2016negotiating}.
Multimodal input interaction~\cite{chokshi2014eplan} and shared interaction~\cite{liu2016shared,liu2017coreach} were found to promote more equitable participation.


Several studies demonstrate that LHRDs are suitable for collaboration, especially as they offer a shared display and support simultaneous user interactions by multiple users~\cite{kister2015bodylenses}.
LHRDs can also accommodate different collaboration styles and the possibility to switch between different styles depending on the task at hand ranging from parallel work to close collaboration~\cite{jakobsen2014up,wallace2016creating}. 
LHRDs come with a variety of interaction capabilities allowing the users to interact up close or from a distance~\cite{beaudouin2012multisurface,jakobsen2014up}.  
Given the large area afforded by the wide screen, seeing more data at the same time turns out to be an effective strategy for collaborative sensemaking.
Previous studies suggest that spatially arranging data enables better intelligence analysis while doing collaborative work in front of LHRDs~\cite{vogt2011co}. 

Mateescu et al.~\cite{mateescu2021collaboration} distinguish three categories of collaborative outcomes: knowledge outcomes, task-related outcomes and social outcomes. 
Collaborative problem solving occurs as a group of people exchange information, ideas or perspectives.
This requires a shared mental model, critical thinking and effective communication when solving the problem.
Users of LHRDs found teamwork to be comfortable and helpful to answer questions~\cite{langner2018multiple}.
User performance gains in terms of task completion time and accuracy have also been reported for pair analytics using an LHRD~~\cite{prouzeau2016evaluating}.

\subsection{Collaborative tools}

Engineering tasks are often performed by teams with diverse experiences and expertise, where teamwork and communication are vital to the project's success. 
Collaborative tools are used to assist stakeholders in the process by facilitating interactions during meetings. 
They can be developed for various platforms e.g., interactive whiteboard~\cite{mangano2014supporting}, interactive tables~\cite{isenberg2011co}, wall-sized displays~\cite{jakobsen2014up}, virtual reality~\cite{he2020collabovr} or multi-surface environments~\cite{mahyar2016ud}. 

There are five categories of tools providing group communication and collaboration:
\begin{enumerate*}[label={\arabic*)}]
	\item Web-based file and document sharing,
	\item real-time conferencing,
	\item non real-time conferencing,
	\item Electronic Meeting Systems (EMS), and
	\item Electronic workspace~\cite{bafoutsou2001comparative}.
\end{enumerate*}
The electronic workspace category is the broadest of all, encompassing a wide range of tools with several functions, and having features of the other four categories too. 
The main concept is to create a shared platform for teams to collaborate and organize their work. 
Groups may store documents and files centrally, collaborate on them, solve problems through discussion. 

Relevant to our work, synchronous collaboration facilitated by large display technologies has been studied in the context of collaborative meetings using electronic workspaces.
The \emph{i-LAND} system~\cite{streitz1999land} offers an interactive landscape including an interactive electronic wall, an interactive electronic table and chairs that are mobile and networked and have built-in interactive devices. 
The \emph{BEACH} software~\cite{streitz1999land} offers the features required for distributing and synchronizing shared information spaces across different devices, and allows users
to move documents or windows between screens.

As touch-enabled mobile devices and large screens have become more affordable a new set of solutions have been based on wall displays.
\emph{WeSpace}~\cite{wigdor2009wespace} is a collaborative tool embedded in a working environment that combines a wall display with a multi-user multi-touch table. 
Based on an ethnographic study of a group of astrophysicists, 
the tool affords to operate several live desktops on multi-touch tabletops and wall displays.
\emph{ReticularSpaces}~\cite{bardram2012reticularspaces} leverages wall displays, tabletops and mobile devices for software development scenarios.
It offers a distributed user interface over different devices applying Activity-Based Computing, 
a context where the user has to establish a precise set of software, hardware, and process requirements.
But the system is not adapted for ad hoc types of collaborative scenarios. 
\emph{CodeSpace}~\cite{bragdon2011code} also uses a multi-surface environment for software development meetings. 
It enables users to transfer information from personal devices to a shared wall display, where everyone can view it. 
The aim is to enable engagement with the surface from a distance by capturing hand motions and mobile devices using the Kinect depth camera. 
Yet again, the system lacks the flexibility required by ad hoc meetings.
ePlan~\cite{chokshi2014eplan} is a tailor-made collaborative tool for emergency response planning. 
It integrates multiple sources of information such as news media, social media, traffic cameras, incident cameras, personnel on the field, and from personnel in the room. 
SAGE2~\cite{marrinan2014sage2,renambot2016sage2} is a generic Web-based collaborative tool supporting group work in front of large shared displays. 
It allows multiple users to display multiple windows on the large display area and interact with them up close or from a distance. 

Like previous systems, WoW supports the Bring-Your-Own-Device (BYOD) paradigm, required in various scenarios where users need to connect to the equipment present in the room.
Although WoW was created with engineering meeting processes in mind, it is flexible and can support generic collaboration activities, which is not the case of ePlan or CodeSpace. 
Among the aforementioned tools, only SAGE2 is Web-based and affords the users an easy access to the environment without any hardware or software restrictions.
WoW differs from SAGE2 as it includes extra capabilities and interactions, described in Section~\ref{sec:features}, with a focus on layout management.

\section{Premilinary study}
\label{sec:prem}

\subsection{Scenario}
\begin{figure}[htb]
  \centering
  \includegraphics[width=\linewidth]{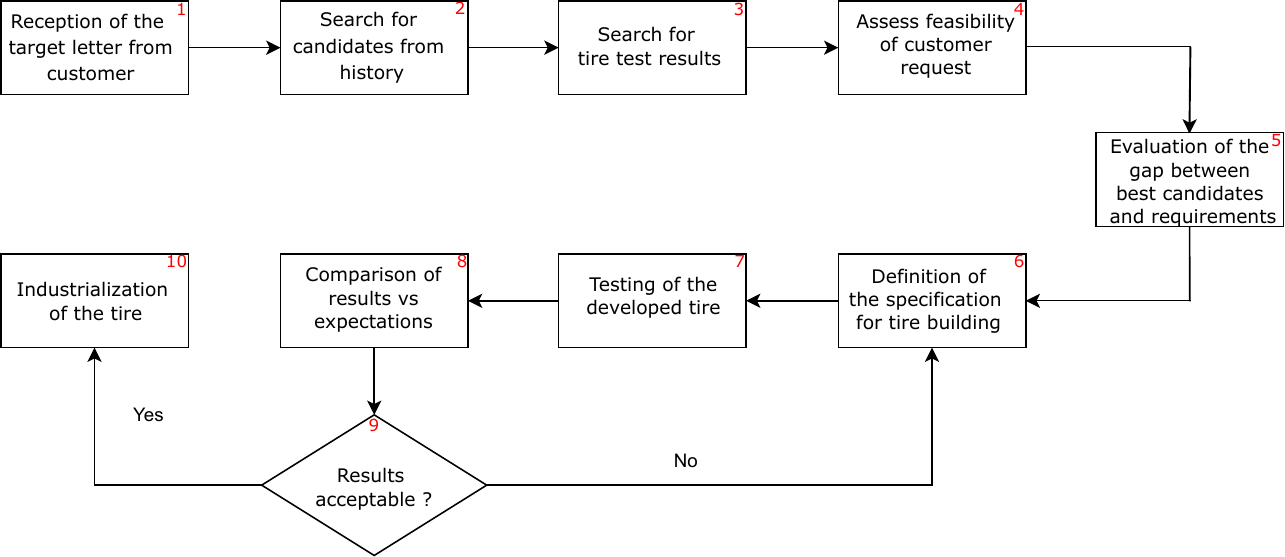}
  \caption{\label{fig:dev_process} Standard steps of a tire development process}
  \Description{wow scenario standard steps}
\end{figure}
Our solution was developed in the context of an industrial R\&D project aiming to improve the efficiency of the design process of pneumatic tires.
Tires are complex composite structures~\cite{Gent2006} that result from an iterative development process to meet the stringent requirements of modern vehicles.
In the rest of this section, by \emph{customer} we mean a car manufacturer, not an end-consumer.
Organized around specific customer accounts with specific requirements,   
these teams include experts in multiple key performance indicators of tires such as comfort, handling, traction and wear.
These engineers need to collaborate, and contribute with their unique expertise so that the final product meets customer requirements.
\autoref{fig:dev_process} shows a typical workflow from the initial customer inquiry up to the final industrialization.

In Step 1, the customer provides a \emph{target letter} describing the requirements for the desired tire. 
For instance, targets are set regarding the rolling resistance, noise levels, lifetime, handling, etc.~\cite{Gent2006}
The customer may also specify one or several existing tires meeting some or most of these criteria, i.e. tire A for rolling resistance and noise and tire B for handling.
Finally the project team has to design a new tire achieving the best of all tires.
Step 2 consists in a high-level look up of tires built in the past that qualify as a good starting point for a new development,  
by leveraging corporate tire databases or drawing on individual knowledge from past projects.
Step 3 targets the specific tire test results for the tires considered in step 2 with respect to the target letter requirements.

While the previous steps are handled by a single engineer,
the full team is consulted for feasibility assessment in Step~4. 
Each expert evaluates the feasibility of the desired tire from her perspective.
As some requirements are contradictory, each design modification is up for negotiation within the full team.
In Step 5 a thorough gap analysis is carried out between the starting tires and the desired tire.
In step 6 the full team works on the definition of a tire specification that would close this gap. 
Multiple solutions are examined in parallel to maximize the chances of success.
Prototype tires are then built.
In Step 7 these tires are tested physically and/or virtually. 
In Step 8 the results are analyzed and a gap analysis is performed. 
The successful design is industrialized in step 10. 
Otherwise, going back to step 6, a new design iteration sets out to identify root causes of the problem and propose new tire designs.

During this process the team meets multiple times to collaborate on problem solving in a creative and innovative manner. 
For every new project, the team defines a brand new product with unique characteristics fitting the car on which it will be mounted. 
In each meeting the participants need to maximize the benefits of teamwork, communicate efficiently, debate, negotiate trade-offs, visualize multi-source data, solve complex problems and innovate.
 
\subsection{Context inquiry study}
Over a 5-month period we sat in several design meetings, observing and taking notes on current practices.
All meetings but one occurred online using MS Teams due to the COVID19 pandemic. 
During the single onsite meeting, a projector was used and driven by the moderator’s laptop. 
In online meetings, participants took turns to share their screen. 

\subsubsection{Preliminary agenda}
Before each meeting, the organizer sends an itemized agenda to the participants,  
sometimes with an estimated duration for each item. 
This helps the audience to anticipate on major discussion items. 
 
\subsubsection{Data and documents}
The documents displayed during the meeting are mostly office documents (.ppt, .doc, .xls, etc.) or PDF files. 
Occasionally the presenter will display other applications. 
Documents may be hosted privately on the participant's laptop or on a shared cloud folder. 
Documents may be static (slide deck) or updated live (spreadsheets, meeting minutes, etc.) and enriched, or corrected based on the discussion. 

\subsubsection{Meeting types and roles}
Despite some variation, meeting discussion items generally fall in two categories:
\paragraph*{Task Management and Planning} Typically, the manager leads this type of meeting. 
This creates awareness and communication on meetings with other team members (project team) and allows for alignment in relation to the initial plan. 
The manager will give out instructions and delegate roles and tasks to participants. 
Usually a spreadsheet (summary file of planning items) or a PDF document (the portfolio and virtual development loop) is used to update the planning in real time based on input from the attendees. 
This document is updated by the person sharing her screen. 

\paragraph*{Technical Discussion}
This type of meeting is led by a Principal Investigator (PI) engineer who discusses technical elements about tires. 
A pre-made slide deck is used to show research results and explain key findings to the audience.
Other participants will verbally ask for a turn to share their own screens and their expertise. 

\subsection{Observed challenges}
\paragraph*{C1. The limited space}\label{par:c1}
The participant's ability to display and share information is limited by the lack of screen real-estate, which adds an overhead on the workflow. 
We observed that the presenter switched documents and applications during her presentation. 
Often taking a screenshot and inserting it in the slide deck was a way to avoid wasting time switching between contents. 
Also, when a participant asked the presenter a question, the latter would shuffle through slides to find the slide related to the question. 
When dealing with several slide decks, the presenter lacks a quick way to view all slides at once. 
More screen real-estate would allow the presenter to show slides along with other files at the same time. 
Complementary views displayed at once would also enhance sensemaking and allow for comparison.

\paragraph*{C2. The need for multiple participants to share content.}\label{par:c2}
In meetings, different experts present their findings in complement to the presenter’s discussion as per the hidden profile paradigm~\cite{stasser1985pooling,lu2012twenty}. 
At times the presenter left a blank slide as a placeholder for a coworker expected to speak.
The latter would have to email her document or upload it online for the presenter to project it. 
The presenter would have to interact with the content e.g., to change slide or scroll down, whilst the coworker explains her findings. 
This interrupts the overall discussions, puts an extra burden on the presenter and potentially frustrates the coworker.
The latter may also seek permission to take the floor and share her screen instead. 
Despite its simplicity, this process impacts the group as the global view is lost and comparisons are less easy to draw when the content is displayed sequentially rather than side by side. 
Participants choose to censor themselves if the displayed content seems more important. 
During one meeting, two participants started discussing an offscreen document to avoid disrupting the presentation.
As the discussion carried on, the presenter stopped sharing her screen, found the said document, took a screenshot and inserted it as a new slide, then shared her screen again. 

\paragraph*{C3. The interaction with the displayed information.}\label{par:c3}
The audience is engaged in the discussion but is unable to interact with or change the content shown by the presenter when the information is stored locally on the presenter’s laptop. 
They are only able to show additional content by sharing their own screen. 

We observed that when a coworker needed to do work in parallel related to the projected content, she would have to interrupt the moderator to ask for details e.g., the location of the document.
For example, when a spreadsheet is displayed and the coworker wants to scroll down to check something, she has to interrupt the moderator. 
We also observed that users had trouble pointing to the displayed content since they did not control the presenter's mouse. 
For instance, when the displayed slide contained several pictures and a participant needed an explanation about one of them. 
It took several attempts before the presenter understood which picture was meant.

\paragraph*{C4. A presenter-audience setting}\label{par:c4}
Since each project has a PI, discussions often adopt a presenter-audience structure.
The presenter moderates the discussion of the results based on a pre-built slide deck.
She is also responsible for updating the shared document with the inputs of participants, even if they can access the document themselves.
We didn't notice close collaboration between users, likely due to a more natural control induced by screen sharing.
Also, the floor is not evenly distributed amongst participants. 
Hence, some of them are more engaged in the discussion than others.

\paragraph*{C5. Development of separate discussions during the meeting}\label{par:c5}
Starting linked but separate discussions is not possible without interrupting the meeting. 
Currently, two experts who want to develop a relevant side topic would have to impose it on the team or leave the room and bring the topic up on their return.

\paragraph*{C6. The inability to capture annotations}\label{par:c6}
During the discussions, if a participant brings up a new idea or suggestion for the displayed content, it is rather difficult to capture it without disrupting the presentation. 
Consequently, capturing annotations during an online meeting is highly inefficient. 
In a regular meeting room, a board can be used, but then no digital record is kept after the meeting.

\paragraph*{C7. Data gathering process}\label{par:c7}
Due to the inherent limited screen space in current meeting setups, generating, combining and editing data is cumbersome.
This often results in action items written in meeting minutes to be performed later on.
In the best case the actions will be performed by a single person, based on her recollection without support from the full team.
Formulating and gathering the new data would be more efficient during the meetings.


\paragraph*{C8. The need to keep track of the session}\label{par:c8}
A lot of content is displayed during meetings: diagrams, presentations, discussion of engineering direction, etc. 
Yet, the fruit of this discussion is never actually recorded.
In the next meeting, the team has to start over and look for the previously shared content. 
This increases cognitive load, appealing to their memory for the same exact content. 
The elements of the discussion are lost, and only textual content can be remembered if written in the meeting notes. 
This usually does not include diagrams or any visual content.

\paragraph{C9. Ergonomics}\label{par:c9}
In regular meeting setups, people are usually sitting at a desk and looking at a screen or a projector.
Interaction and collaboration is limited to verbal inputs while seated.
Breaking out into groups to start a related side discussion, to solve a problem in parallel, like in workshops, is impossible.
People are stuck in their place and cannot start a collaboration using a part of the available screen space.

\subsection{Design goals}
From the observed challenges, we identified the following design goals.
\paragraph{D1. Collaborate on problem solving and share insights}\label{par:large_display}
We believe that tire engineering meetings can benefit from the increased display size and resolution of LHRDs  since they support the display of multiple sources of information at once and improve sensemaking in collaborative problem solving contexts~(\hyperref[par:c1]{C1}). 
Users can reach more observations and broader insights with LHRDs than with regular displays~\cite{reda2015effects}. 
Studies showed that LHRDs have the potential to improve collaborative problem solving (see Section \ref{ssec:colabLHRD}). 

\paragraph*{D2. Interact from a distance or up-close}\label{par:interact}
The selection of input interaction can influence many dimensions of group processes: awareness, interference and collisions, use of space, equality of participation~\cite{jakobsen2016negotiating}.
Our design goal is to enable smooth transitions between working with the wall using direct touch and interacting from a distance using personal devices~(\hyperref[par:c3]{C3}).
This promotes physical navigation. 
Users can get an overview of a complex visualization by stepping back, and get a closer look at details by moving closer to the screen.

\paragraph*{D3. Integrate multiple sources of data}\label{par:multi_source}
In the observed tire engineering meetings, decisions are usually made by collecting and judging multi-source information. 
A larger display can fit multiple views at once and ease access to any of them, which can improve the workflow~\cite{bi2009comparing}(\hyperref[par:c2]{C2}). 
We aim to support all types of data needed during these meetings to avoid any constraint when using our solution. 

\paragraph*{D4. Support participation equity and different collaboration styles}\label{par:collab}
Our solution also needs to promote an equitable communication within an interdisciplinary team.
Seeing other partners' actions could promote higher participation~\cite{feibush2000visualization}. 
We aim to provide users with different modalities to interact simultaneously with the displayed content so that all users have an equal opportunity to control the discussion at all times, preventing a single user from monopolizing the system and thus the conversation. 
We want to avoid presenter-audience settings~(\hyperref[par:c4]{C4}). 
Our solution should accommodate different collaboration styles, ranging from parallel work, to discussions, to close collaboration~(\hyperref[par:c5]{C5}).

\paragraph*{D5. Spend less time on configuration}\label{par:time_config}
During meetings, participants should focus on collaborating on problem solving, not on software/layout configuration.
To this end, we will provide means to do preparatory work prior to the meeting~(\hyperref[par:c7]{C7}).
Session content will also be saved and retrieved at the start of the next meeting~(\hyperref[par:c8]{C8}).  
We will also integrate useful features e.g., annotations (\hyperref[par:c6]{C6}) and provide appropriate interactions for LHRDs e.g., to limit unnecessary movements in front of the LHRD~(\hyperref[par:c9]{C9}).


\section{System description}
\label{sec:sysdesc}

\subsection{System overview}

\begin{figure}[htb]
  \centering
  \includegraphics[width=\linewidth]{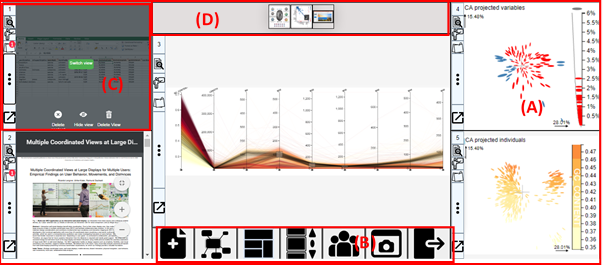}
  \caption{\label{fig:wow_overview} Overview of the graphical user interface of WoW.}
  \Description{Overview of the graphical user interface of WoW.} 
  \label{fig:WoW_Overview}
\end{figure}

WoW is a multi-user Web application for collocated and remote collaborative meetings and workshops in multi-surface environments.
Users can just use a Web browser to connect to it from their own devices, simultaneously upload documents and show them on the LHRD.
WoW is intended for the visualization and cross-analysis of multiple data sources taking advantage of the increased display space and resolution.
The user interface of WoW supports multimodal interactions, including direct touch and interacting from a distance using personal devices.
\autoref{fig:wow_overview} shows the main components of the graphical user interface of WoW.

\paragraph*{(A) Multiple viewports}
The available space is partitioned into contiguous rectangular viewports of variable size where content can be loaded and displayed side by side.
This caters for the need to analyze multi-source information. 

\paragraph*{(B) Quick access toolbar}
The quick access toolbar is an area located at the bottom center of the screen containing a set of frequently used commands that are not reliant on a certain view.
The central location is chosen as users tend to gravitate toward the center of LHRDs as the main interaction region~\cite{bi2009comparing} and because the top is not reachable by all.

\paragraph*{(C) View manipulation layer}
During a workshop, users may need to modify the content of existing viewports.
The use of small icons, like in the task bar of the Windows operating system,
is time-consuming and cumbersome~\cite{robertson2005large}.
The view manipulation layer expedites basic operations such as swapping and editing views by overlaying them with broad target regions.
This lowers the level of accuracy needed for user interaction, frees up space and prevents the interface from becoming cluttered with too many icons displayed for each view.
This design choice is inspired by the work of Rooney et al.~\cite{rooney2012improving} who designed an interaction layer for resizing windows in wall-sized display operating system.

\paragraph*{(D) Hidden views stack}
Because all documents are not required at all times, a stack of hidden documents is kept at the top center.
These documents can be brought back on display as soon as they are relevant to the discussion.
This preserves the dicussion context while freeing up space for active content.


\subsection{System architecture and implementation details}
\begin{figure}[htb]
  \centering
  \includegraphics[width=0.65\linewidth]{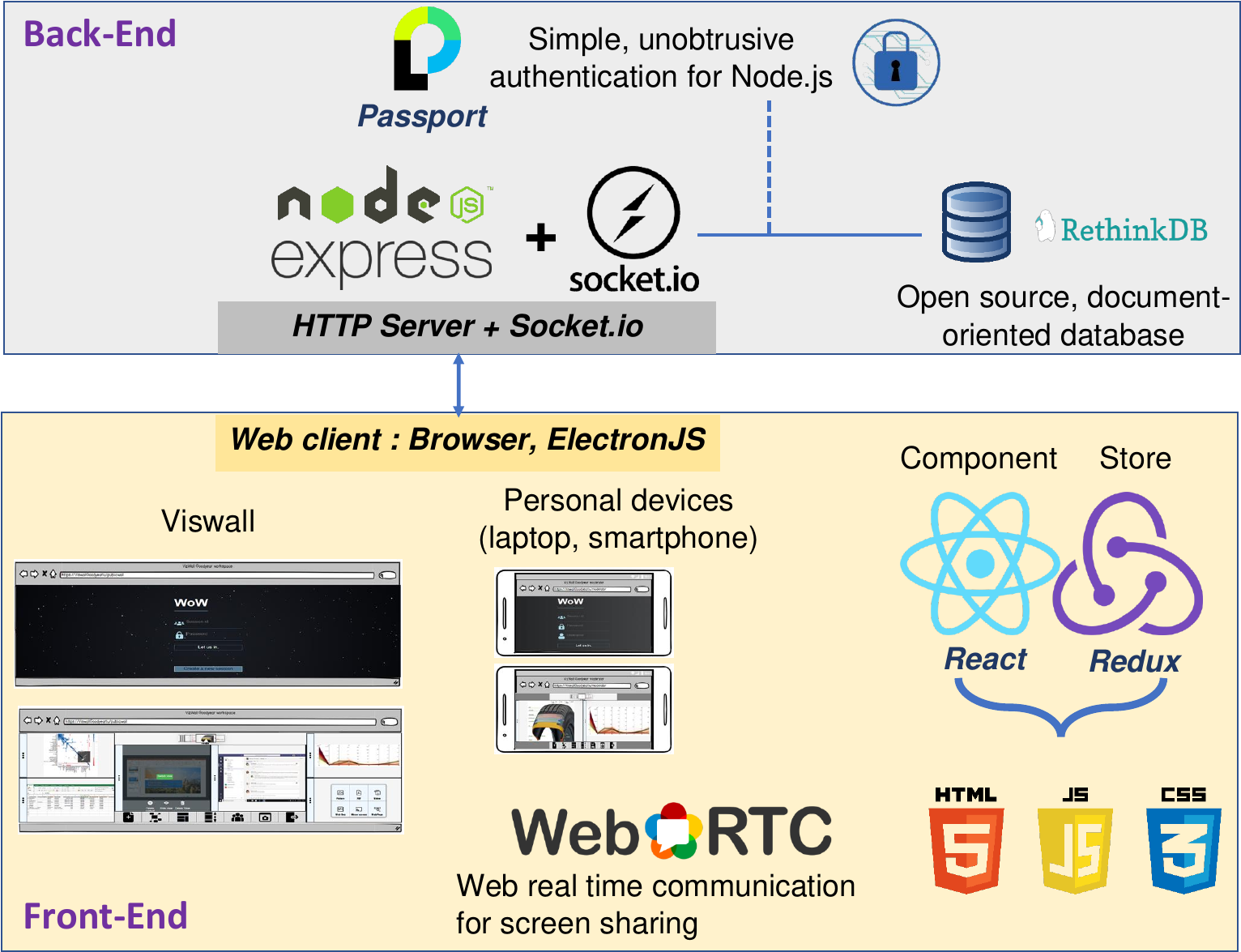}
  \caption{\label{fig:Architecture_WoW} An overview of the software architecture of WoW.}
  \Description{WoW Architecture}
\end{figure}

\autoref{fig:Architecture_WoW} shows a diagram of the architecture of WoW.
It comprises a Web server based on the Node.js technology, in conjunction with Express, a popular JavaScript Web application framework.
The Javascript frontend is developed as a single-page application based on React.js and Redux according to the model-view-controller pattern.
The responsive user interface makes WoW accessible from a variety of devices, including a wall-sized display, laptops, tablets, and smartphones.
Various data containers (for static file formats and Web pages) are created with HTML and JavaScript.
We use \emph{WebRTC} to support screen sharing within the viewports of WoW.
Web sockets ensure real-time communication between the clients (wall-sized display, tabletop, or user device) and the server and to keep all clients in sync.
Session content is persisted in real-time using RethinkDB database.
Required in corporate environments, user authentication and session management is implemented using \emph{Passport}, an authentication middleware for Node.js.

Advanced capabilities include controlling the mouse pointer of the LHRD, and displaying third-party Web pages in the viewports, from client devices.
To avoid clickjacking attacks, these features are only available when WoW is packaged as an Electron application.

\subsection{Features and operations}
\label{sec:features}
The WOW platform offers the following features:

\subsubsection{Flexible layout management}
A large number of views can quickly complexify the graphical interface.
Hence, layout management plays an key role to ease work with multiples views and improve the user's workflow.
WoW lets users easily add, delete, re-arrange views with little interaction, time and effort during the meeting, as detailed below.

\begin{figure}[htb]
  \centering
  \includegraphics[width=\linewidth]{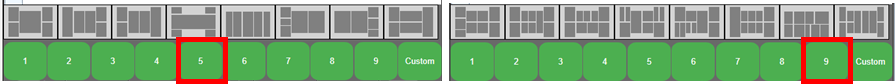}
  \caption{\label{fig:wow_selectPreset} Available presets for a 5-view layouts (left) and 9-view layouts (right).}
  \Description{Available presets for a 5-view layouts (left) and 9-view layouts.}
\end{figure}

\paragraph*{Select a preset layout}
Layout presets are the basic building blocks to lay out the views needed for a work session.
To save time, WoW comes with predefined layouts accessible from the Quick access toolbar. 
The user can select the desired number of views and one of the available arrangements, as in~\autoref{fig:wow_selectPreset}.
The layout can later be edited during the session.

\begin{figure}[htb]
  \centering
  \includegraphics[width=0.95\linewidth]{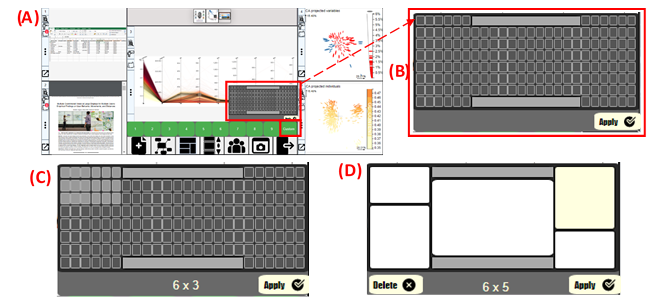}
  \caption{\label{fig:wow_customLayout} After selecting the custom layout option (A), an empty preview grid is shown (closeup in (B)). The first view is specified (C) followed by the other views and the final result is previewed (D) before applying it.}
  \Description{Steps of building a custom 5-view layout. The first view is specified (bottom left image) and the preview of the final result (bottom right).}
\end{figure}

\paragraph*{Build a custom layout}
Custom layout can also be built from scratch.
The wall layout is organized as a rectangular grid.
Viewports are specified one at a time by providing the dimensions of the desired rectangle, similar to common table creation wizards, as shown in~\autoref{fig:wow_customLayout}.
A preview of the result is shown before applying any changes.

\begin{figure}[htb]
  \centering
  \includegraphics[width=\linewidth]{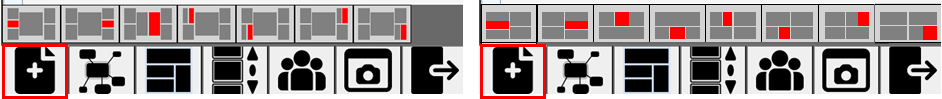}
  \caption{\label{fig:wow_addNewView} Placement options for inserting a $6^{th}$ in a 5-view layout (left), or a $5^{th}$ view in an existing $2 \times 2$ grid (right).}
  \Description{Add a new view}
  \label{fig:addView}
\end{figure}

\paragraph*{Insert a new view}
After a layout is created, users can still insert new views into it (\autoref{fig:addView}).
Simple heuristics are used to place the new view, such as
creating a space between two views by reducing their space horizontally or vertically, or halving an existing view, or placing the view in available empty space in the layout.

\begin{figure}[htb]
  \centering
  \includegraphics[width=\linewidth]{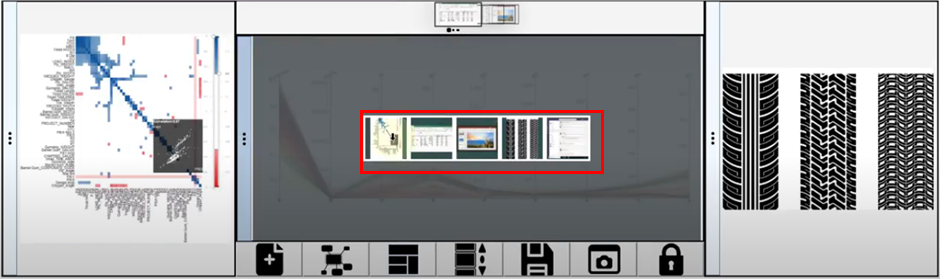}
  \caption{\label{fig:wow_switchViews} Swapping views. Thumbnails ease the choice of the target location instead of drag-and-drop interactions.}
  \Description{Swapping views}
\end{figure}

\paragraph*{Swap two views}
Views get added and deleted during use as discussion and workflow unfold.
Unlike windows in regular desktop environments, relocating a view on a wall-sized display may entail a lot of walking. 
Also, prolonged touch input could be frustrating, unpleasant to users and accidentally interrupted leading to unexpected results. 
Owing to its grid layout, WoW provides view swapping as an alternative to drag-and-drop interaction.
After selecting the view to be moved, the user is shown the thumbnails of all visible and hidden views to swap with, as in~\autoref{fig:wow_switchViews}.

\begin{figure}[t]
  \centering
  \includegraphics[width=\linewidth]{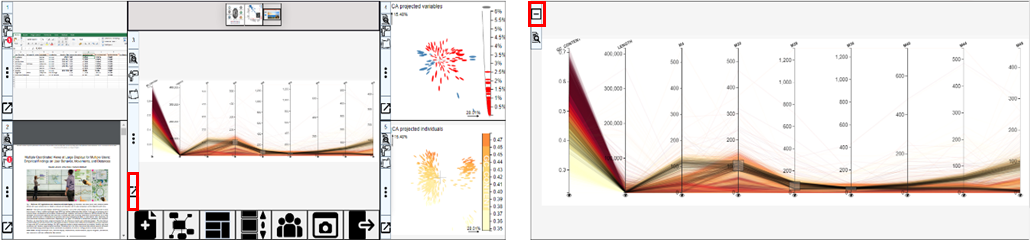}
  \caption{\label{fig:wow_fullScreenView} Example of maximizing a parallel coordinates plot.}
  \Description{Expand a view as a full screen}
\end{figure}

\paragraph*{Maximize a view}
Blowing up a view to occupy the entire display space may be helpful. 
On LHRDs this supports the analysis of ultra-high resolution content like gigapixel images or hires visualizations (\autoref{fig:wow_fullScreenView}).
On personal devices a view of interest gets the entire pixel budget, instead of replicating the entire wall display with tiny views.

\paragraph*{Work with multiple virtual walls}
Like virtual desktops, virtual walls provide a way to organize content into multiple spaces and switch between them easily e.g., to support storytelling or discuss several design scenarios.
Users could also divide the views by activity, to better focus on that activity.
From the quick access bar, the user gets a live thumbnail preview of the virtual walls to easily switch between them (\autoref{fig:wow_multipleWalls}).
Each virtual wall may have its own layout.

\subsubsection{Support for several data formats}
WoW can render various file formats including static files (PDF, images and videos) and Web links pointing to documents stored in the cloud e.g., office documents.
Screen sharing affords users to show any application running on their personal devices or their entire screen, thus alleviating any file format limitations.
A single user can also share multiple windows/screens at once.

\subsubsection{Direct and distant interaction}
Concurrent interaction with multiple views of the shared display creates new and interesting ways to collaborate e.g., to take over certain actions, present, and share results simultaneously.
Coworkers can use their own smartphones, tablets and laptops to interact with the content displayed on the wall. 
Concurrent interaction with the shared display is supported in three ways:
\begin{enumerate*}
  \item direct touch interaction on the wall;
  \item remote interaction on the content displayed on a personal devices. These interactions are synchronized across all wall and personal displays;
  \item remote interaction using a virtual mouse cursor shown on the wall which acts on the content of the wall display.
\end{enumerate*}
Each virtual cursor is labeled after its owner to increase mutual awareness.
These cursors are handy for remote presentation purposes.
Visual feedback is displayed if the user triggers any pointer action.

\subsubsection{Note taking}
\begin{figure}
\centering
\begin{minipage}{.45\textwidth}
  \centering
  \includegraphics[height=4.25cm]{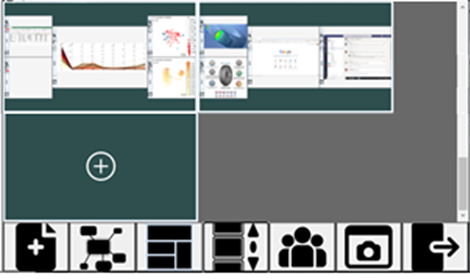}
  \captionof{figure}{\label{fig:wow_multipleWalls} An example with two virtual walls to support complex meeting workflows and storytelling.}
\end{minipage}\qquad
\begin{minipage}{.45\textwidth}
  \centering
  \includegraphics[height=4.25cm]{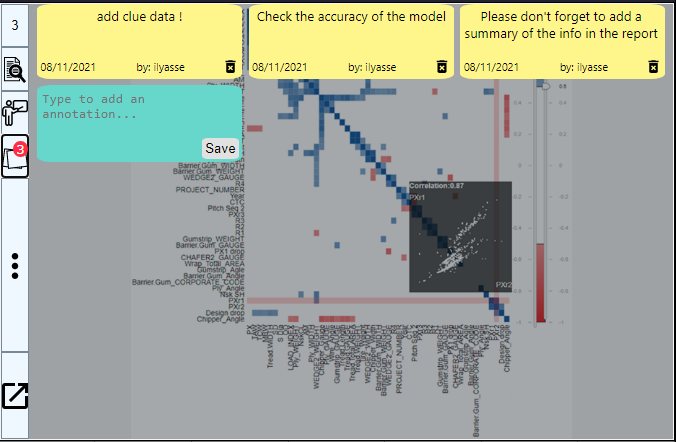}
  \captionof{figure}{\label{fig:wow_noteTaking} Example of notes taken concerning a correlation heatmap.}
\end{minipage}
\end{figure}

To improve the organization and usage of notes, WoW supports the ability to add notes to any view via the interaction layer (see \autoref{fig:wow_noteTaking}).
Notes can be taken from any connected device. 
Contextual information like the author and related view help the post-hoc comprehension of notes.

\subsubsection{Session management and storage}
To support activities running on extended periods, WoW saves session content and can respawn it later on, including view layout.
Meetings can then be resumed quickly.
In a way, this mechanism  also keeps a trace of how and why past decisions were made. 

\section{Study}
\label{sec:study}

\subsection{Method}
To evaluate the usefulness and usability of WoW for pneumatic tire design workflows, we proposed to two distinct teams of tire engineers to use WoW to conduct real design meetings as part of their ongoing projects.
We conducted two design sessions aiming to specify new tire builds (\autoref{fig:dev_process} - step 8): 
\begin{enumerate}
    \item The first session lasted three hours. We used this occasion to run a usability study of WoW with five onsite participants and five remote participants followed by a semi-structured focus group to get various viewpoints that might help us understand the usage and improve our solution. 
    \item The second session lasted a whole working day, interrupted by a lunch break and a coffee break. We used this occasion to run a usability study of WoW with 10 onsite participants and 2 remote participants, followed by SUS~\cite{brooke2013sus} and NASA-TLX~\cite{hart2006nasa} questionnaires and open questions to collect subjective quantitative data and assess how it aligned with qualitative data.
\end{enumerate}

\subsection{Participants}
All study participants are with the Goodyear Innovation Center Luxembourg. 
Their main role is to develop new tires for consumer vehicles, especially for major car manufacturers whose identity is withheld. 
They met to discuss and tune different tire specifications built for an upcoming car make according to customer requirements (\autoref{fig:dev_process} - step 1).
The project manager led the meeting together with a principal investigator for this particular tire development. 
The other participants were experts in specific fields like acoustics, vehicle handling, tire wear etc. 
As some tire performance goals conflict, expert feedback and collaboration was extremely important for the success of the project. 
Five attendees were collocated in front of the LHRD, and five participated remotely through a conferencing software, including one female.
Their ages ranged from 25 to 60 years.
Before they met, participants had prepared material to share, 
including office documents, in-house applications or commercial signal processing software.

\subsection{Setup}
The experiment took place in a multi-surface environment comprising a touch-enabled wall-sized display, a 55-inch horizontal touch-enabled tabletop and personal laptops (see \autoref{fig:teaser}).
The wall-sized display serves as an information radiator aggregating data from many sources and making it available to the public.  
The tabletop serves as a semi-public collaboration area.
Laptops serve as personal private workspaces from which users can share private data and interact with the platform.
The wall-sized display comprises 24 high-definition tiles arranged in a $2\times 12$ grid ($\approx 7m\times 2m$), with a total resolution of $12690\times 3840$ pixels.
The interactive tabletop is placed in the middle of the room and is connected to the active session displayed on the wall-sized display.
It allows several coworkers to use WoW concurrently.

Each participant in the room brought his own laptop and signed onto WoW to interact with the wall and share content like local files, files from the cloud, or screen sharing. 
Remote participants were signed onto WoW too. 
In addition, they were connected via Microsoft Teams to engage in the discussion through a dedicated laptop connected to a videoconferencing device for big rooms with a wideband loudspeaker and an omnidirectional microphone. 
A dedicated audio system was set up for the second meeting to improve the audio experience of remote participants. 
The web camera of the laptop was directed towards the wall so that remote participants could capture the context of the discussions and understand what was going on in the room.

\section{Findings}
\label{sec:findings}

\subsection{Observational findings}
\paragraph*{Multiple views and the use of space}
During meetings, various contents (\hyperref[par:multi_source]{D3}) were shared in multiple views, ranging from three to nine views at once.
We took note of two different uses: displaying several views in the chronological order of discussion or displaying several views that will be used concurrently and represent additional information (\hyperref[par:large_display]{D1}). 
Users have used the different layout functionalities to organize the content to their liking. 
They created a layout and adjusted it on the fly, adding new views, updating the content of existing views or deleting views that were no longer needed, and swapped views to place two views side by side when the discussion concerned them both (\hyperref[par:collab]{D4}).

During the second session, we also noted that one of the users frequently switched to full screen mode on the wall display when presenting. 
This is probably due to an acquired habit of displaying a document on a projector in a presenter-audience mode.
The shared material was limited to SharePoint links (to share slide decks or spreadsheets) or screen sharing from a laptop to show third-party software e.g., Tableau or Matlab. 
Static files were not used because project files are hosted in the corporate cloud and must be synced across all participants. 

\paragraph*{Interaction modality and device usage}
The coworkers focused on the wall display. 
We didn't see any users undertaking non-meeting work on their own device during the meeting (\hyperref[par:collab]{D4}). 
Personal devices were used to share content or update a document in real-time while screen sharing the personal device with others on the wall for meeting-related content. 
Layout management was mostly carried out by touching the wall directly (\hyperref[par:interact]{D2}). 
The presenter has notably made use of direct touch on the wall, except when using the screen sharing mode from her personal device.

\paragraph*{Teamwork and communication}
The coworkers were more dynamic and engaged in the discussion than during their conventional meetings (\hyperref[par:collab]{D4}). 
They asked more questions, relied on multiple views to better convey ideas and contributed more to solving the problem at hand. 
They also engaged in parallel discussions. 
In mixed-presence meetings, awareness strategies are needed to help transmit the context to remote users and better involve them in the discussion as much as possible. 
Other options include having larger hires displays available to remote users.

\paragraph*{Movement and physical space}
Although the sessions ran for a long time, users kept standing and walking in front of the wall display. 
Some participants sat on chairs from time to time. 
More walking occurred when interacting with contents laid out far apart on the wall.
Presenters and the discussion moderator tended to stay close to the wall and walk more than others.
We noted more movement in the first session than in the second, which may have two explanations. 
First, the nature of the work done differed in the two sessions.
The first sessions dealt with a project in its early design iterations and required more user engagement, while the project of the second session had been through more iterations and could use more storytelling.
Second, the number of coworkers in the room may affect their movements. 
The group was larger in the second session, and people refrained from moving to avoid occluding the wall display.


\subsection{Qualitative results}
\begin{table}[htbp]
  \centering
  \resizebox*{!}{0.97\textheight}{
\begin{tabular}{|p{17.5cm}|p{5.5cm}|p{2.5cm}|} 

\hline 
\rowcolor{Gray}
\Large\textbf{Meaning Units (Condensations)} & \Large\textbf{Codes } & \Large\textbf{Categories} \\ \hline 
\multicolumn{3}{|p{14cm}|}{\Large\textbf{\textsc{Theme~1: Comparing The Solution To The Standard Meeting Setting}}} \\ \hline 
\Large``More content on same screen'' {}\newline ``Opportunity to have more things on screen''\newline ``Possibility to display huge amount of data at the same time''\newline ``Ability to project multiple documents''\newline  ``Multiple document displayed at same time''\newline ``More data can be reviewed at same time'' &\Large More data at once & \multirow{9}{2.2cm}{\Large\textbf{Benefits of LHRD }}  \\ \cline{1-2} 
\Large``Good to be able to show multiple sources'' &\Large Multi-source data &  \\ \cline{1-2} 
\Large``Good overview of several documents''\newline ``The multi display helps to keep an eye on several objects / screens at the same time'' &\Large Multiple documents, overview  &  \\ \hline 
\Large``All experts in the same room working collaboratively''.  &\Large Collaboration work & \multirow{18}{2.2cm}{\Large\textbf{Impact on collaborative problem solving}} \\  \cline{1-2} 
\Large``Parallel workflow is possible'' {}\newline ``With the wall, the discussion is more multi-objective / parallel instead of mono objective/ sequential'' &\Large Parallel workflow &  \\ \cline{1-2}
\Large``All working on the same room'' & \Large Co-located work &  \\ \cline{1-2} 
\Large``All people look on one big screen and not on their own separate screens''\newline ``All looking at the same screen'' &\Large Shared display &  \\ \cline{1-2}
\Large``All people are more awake and contribute'' {} \newline ``Like in workshops, all people awake'' &\Large Awaking / more engagement &  \\ \cline{1-2}
\Large``People are more engaged in the discussions'' {}\newline ``People standing making more engaged''\newline ``More involvement from people'' &\Large More engagement &  \\ \cline{1-2}
\Large``Everybody can break out, initiate separate discussions without bothering other attendees'' {} &\Large Separate discussions, parallel workflow &  \\ \cline{1-2}
\Large``More interactive discussions''\newline ``More interaction/ discussion, not just listening to one person presenting'' &\Large More discussions &  \\ \cline{1-2} 
\Large``The meeting becomes a real workshop''\newline ``Different way of working'' &\Large Group dynamics &  \\ \hline 
\Large``Less ping-pong between data as all are displayed on the wall'' &\Large Less display switching / effort & \multirow{20}{2.2cm}{\Large\textbf{Impact on work efficiency and outcome }}  \\ \cline{1-2} 
\Large``With wall there is less time lost to change presentations''\newline ``Avoid a loss of time due to screen change''\newline ``Less time loss to retrieve a document / presentation'' &\Large Less display switching / time  &  \\ \cline{1-2}
\Large``It looks more interactive'' &\Large More interactions / discussions &  \\ \cline{1-2}
\Large``Very efficient usage as you can look on many presentations and no need to switch''\newline ``More efficiency in discussion in front of Viswall'' &\Large More efficiency  &  \\ \cline{1-2} 
\Large``Possibility to switch quickly to different content'' &\Large More efficiency / Flexibility  &  \\ \cline{1-2} 
\Large``Flexibility in making views, many different ways to share and operate''\newline ``Dynamic of view switching to support ongoing discussion'' &\Large Flexibility  &  \\ \cline{1-2} 
\Large``Attendees don't need to remember what was on screen before as more content can be displayed'' {} &\Large Memory &  \\ \cline{1-2} 
\Large``Good for decision making'' {}  &\Large Better Decision- making &  \\ \cline{1-2} 
\Large``We reach a more informed decision based on new knowledge shared among us'' &\Large Better Decision-making / Collaborative sensemaking  &  \\ \hline 
\multicolumn{3}{|p{18cm}|}{\Large\textbf{\textsc{Theme~2: Scenarios When Users Believe The Solution Facilitates The Meeting Wokflow}}} \\ \hline 
\Large``Decision, data analysis, collaboration meetings'' &\Large Context: collaborative decision making & \multirow{9}{2.2cm}{\Large\textbf{Opportunities and application scenarios}}  \\ \cline{1-2} 
\Large``Decide on design in case of trade-offs'' &\Large Context: trade-offs cases  &  \\ \cline{1-2} 
\Large``In large workshops as we had today / use data from different sources and share to larger audience.'' &\Large Context: multiple source data  &  \\ \cline{1-2}
\Large``Many people involved'' &\Large Context: teamwork  &  \\ \cline{1-2} 
\Large``Fully dedicated to solving the problem'' {} &\Large Context: complex problem solving  &  \\ \cline{1-2} 
\Large``During workshops where a lot of presentations / data are shared.''\newline ``much data to be shown'' &\Large Context: multiple documents  &  \\ \hline 
\multicolumn{3}{|p{19cm}|}{\Large\textbf{\textsc{Theme~3: User Suggestions And Issues Related To User Experience, Interface, And Configurations}}} \\ \hline 
\Large``Be able to display MATLAB / X3D Catia on Viswall'' &\Large External app integration & \multirow{10}{2.2cm}{\Large\textbf{Interface enhancements }} \\ \cline{1-2} 
\Large``Hide menu to maximize space on screen'' &\Large Improve data-ink ratio &  \\ \cline{1-2} 
\Large``Possibility to zoom in on part on a section for quick review.''\newline ``A zoom mode.'' &\Large Zoom &  \\ \cline{1-2}
\Large``Directly compute and sketch in the presentation on the wall'' &\Large Realtime sketching &  \\ \cline{1-2} 
\Large``Improve quality \& Involvement of people connected remotely''\newline ``As a remote user having a fixed camera could be good to follow the discussion''\newline ``Have remote people on the screen''\newline ``Not disruptive for people who are physically in but can be disruptive for remote people'' &\Large Telepresence &  \\ \hline 
\Large``SAS JMP graphs not same on PC vs screen'' &\Large Color settings  & \multirow{8}{2.2cm}{\Large\textbf{Room and wall set-up}} \\ \cline{1-2}
\Large``The lights on the Viswall are too powerful'' \newline ``Eye- tiredness of high-luminosity of screen.'' &\Large Light settings &  \\ \cline{1-2}
\Large``Tactile feedback not good as my iPhone''  &\Large Tactile feedback &  \\ \cline{1-2}
\Large``Organize a little better space in front of Viswall (more chairs and table)''\newline ``Bigger room and more interactive desks'' &\Large Room / furniture layout  &  \\ \cline{1-2} 
\Large``Control panel.''\newline ``A control panel (was not used during the session)'' &\Large Tabletop control panel &  \\ \hline 
\Large``Refresh rate.''\newline ``Refresh of content / sometimes discounted from server'' &\Large Server timeout & \multirow{7}{2.2cm}{\Large\textbf{Issues and glitches}} \\ \cline{1-2} 
\Large``Prepare a 5 minutes video with instructions for all participants (like a pre-briefing) before the working session''\newline ``A small training or manual received before the 1st use.''\newline ``Provide guidelines before starting the session''\newline ``Maybe provide a solution for beginner'' &\Large Training  &  \\ \hline 
\end{tabular}}

  \caption{\label{table:ContentAnalysis} Content analysis: Step1) Identifying and condensig meaning units from focus group and questionnaires. Step 2) Formulating codes with the condensed meaning units. Step 3) Creating categories for the codes. Step 4) Creating themes for the categories.}
  \Description{Content analysis}
\end{table}

We used a four-step method to analyze the qualitative data collected from the focus group of the first session and the open-ended questions of the questionnaire in the second session~\cite{erlingsson2017hands} (See \autoref{table:ContentAnalysis}).
The analysis was carried out manually. 
After highlighting important concepts and categorizing the information, the text of user feedback was divided into 67 semantic units. 
Each unit focused on a topic, which was in-turn condensed/divided into further sub-units or phrases.
Notably, the condensed units, however short, retained the core meaning of the text. 
Hence, condensed units with similar core-meaning were grouped. 
After condensing the data, two independent coders coded the semantic units.
A third coder was involved to help resolve disagreements. 
Then, the three coders assigned the agreed codes to categories and themes.
Finally, three themes were created to provide a more precise answer to the main research questions. 

Findings from our content analysis suggest that participants felt that WoW could improve workflow efficiency, collaboration, engagement and decision-making. 
The interface and interaction design, which provided flexibility to manipulate the different views, was also highly appreciated and deemed as one of the primary strengths of the application. 
Also, some participants stated that WoW had the potential to turn design meetings into real workshops.

Several users pointed out the added value of displaying several contents from multiple sources at once. 
Some concerns regarding user experience, interface design, and the configuration were also expressed. 
Suggestions included hiding the menus when they were not needed; the ability to zoom within each view; support for rendering more data types on the wall like Matlab documents, and the ability to sketch CAD models in real-time. 
Other improvement ideas included a better support for remote participants in mixed-presence meetings, as well as the ability to see them on the wall display. 
Concerning the room configuration, participants wanted more furniture to be more comfortably seated during long meetings. 
Room lighting and the luminosity of the wall display also needed to be adjusted to avoid eyestrain.

In sum, participants consistently expressed enthusiasm about the usefulness, usability, and impact of WoW as a viable tool to support collaborative design meetings. 
They also identified several limitations and improvement directions.

\begin{figure}[t]
  \centering
  \includegraphics[width=\linewidth]{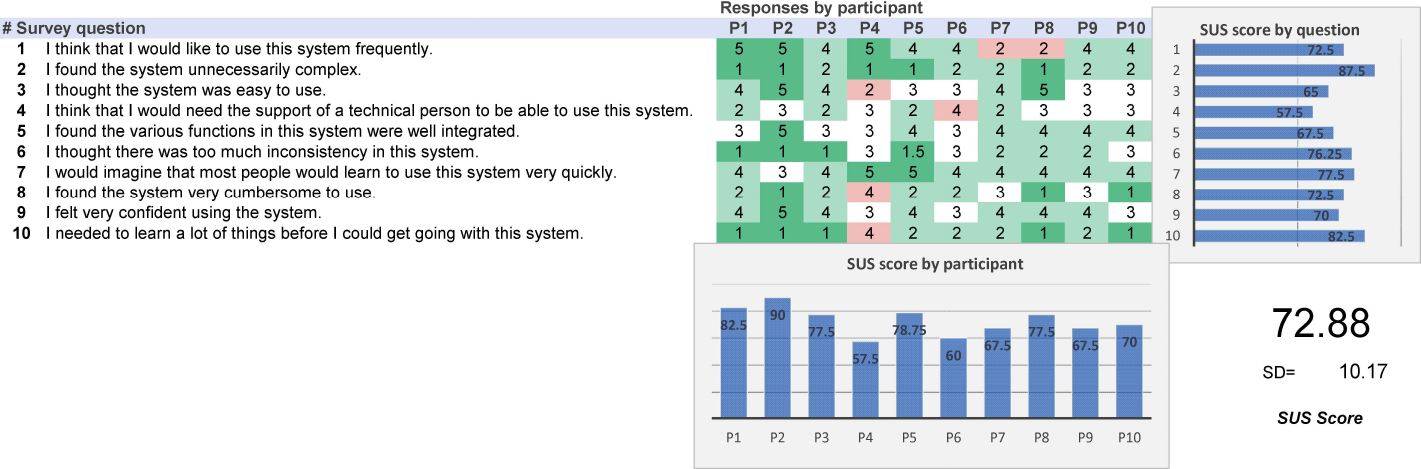}
  \caption{\label{fig:SUS_score} SUS usability score}
  \Description{SUS usability score}
\end{figure}

\begin{figure}[t]
  \centering
  \includegraphics[width=10cm]{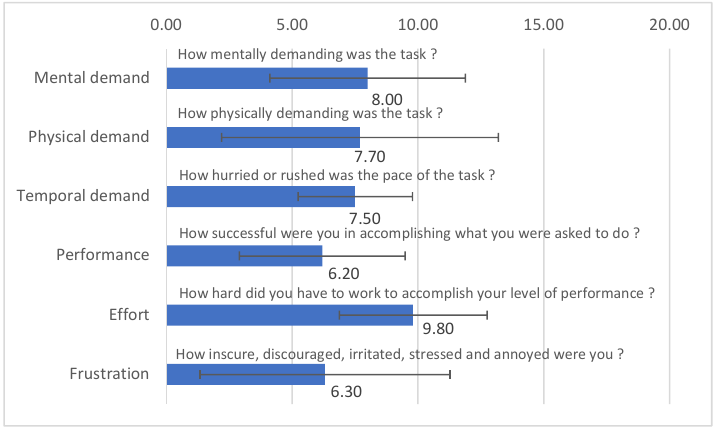}
  \caption{\label{fig:NASA_TLX_score} Average NASA TLX score}
  \Description{Average NASA TLX score. In all cases, a lower score
  is better (e.g., for performance, a higher score indicates a perfect performance}
\end{figure}

\subsection{Quantitative results}\label{sec:quantitative_results}

The system usability scale questionnaire in the second session yielded an average score of 72.88 (SD=10.17), which is a very good measure of satisfaction~\cite{bangor2008empirical} (see \autoref{fig:SUS_score}).
The perceived workload from the NASA TLX~\cite{hart2006nasa} test was also good, better than average for all metrics, especially performance and frustration (see \autoref{fig:NASA_TLX_score}). 
One participant linked the physical demand to the fact that people were standing during the meeting. 
Although participants did experience limited demand overall, the effort was rated slightly higher, at 9.80.

\section{Discussion}
\label{sec:disc}

In this section, we focus more on the analysis of the qualitative findings, with some supporting comments from the quantitative data from the usability and cognitive load results.

The qualitative analysis has clearly shown that the flexibility of view layout in WoW made it an effective means to analyze more content simultaneously, which is   also supported by the good outcomes of the usability and NASA tests.
Users were positively impressed by the support for collaborative problem solving (Section~\ref{sec:CPS}). 
They felt that having so much information readily displayed greatly improved the workflow and meeting efficiency, and supported decision-making. 
They also felt that accessing multiple views at once would reduce data exploration time and allow the team to find actionable ideas promptly. They also found WoW relevant in problem-solving and decision-making situations, as well as in use cases where trade-offs needed to be debated.
Lagner et al.~\cite{langner2018multiple} displayed as many as 47 views at the same time on an LHRD.
In settings smilar to tire design meetings, the typical number of views is yet unclear.
Further user studies are needed to evaluate the complexity of coordinated multiple view systems on LHRD for industrial usage.

Although participants deemed the interface easy to use and learn in the usability testing, they also made some suggestions, such as the necessity to receive a tutorial to use the system properly. 
This aligns with the mixed feeling that came out of the question ``I think that I would need the support of a technical person to be able to use this system'' which received a score of 57.5/100. 
Yet, they thought WoW had a shallow learning curve, as ``Quick system usage'' scored 76.5/100. 
Users also requested features in relation to improving the experience of remote users. 
This shows that supporting remote users in distant team collaboration environments constitutes an important improvement direction.
Besides being a necessity in multinational companies with many different sites, this requirement will probably be emphasized with the increasing use of home office after the COVID19 pandemic.
One possibility to improve telepresence and remote collaboration across several sites consists in displaying the video of remote users on the LHRD taking into account different placement strategies for the video according to the respective position of the local and remote user as suggested by the study of Avellino et al.~\cite{avellino2017camray}.

All in all, our quantitative and qualitative findings provide strong evidence for the potential value of WoW as a tool for more engaged collaborative meetings. 
Thanks to the multiple coordinated views and the large physical space of the wall display, engineers may use WoW to support decision-making and optimize collaborative design workflows.
\section{Conclusion and future work}
\label{sec:concl}
Displaying more content at once promises to give engineers a convenient overview of several complementary documents. 
This alleviates cognitive challenges that engineers face while working with large, multi-source data. 
Next generation of collaborative environments, such as the proposed software, will help engineers be more effective at synthesizing and retaining contextual information during their exploratory data analysis, ultimately leading to better decision making.
This work gives some important directions regarding the potential of specialized software applications for wall displays. 
WoW also opens other research opportunities to overcome the hurdles of an array of encumbered interactions with large wall displays and their use in effective collaborative meetings and workshops. 

The present study has two primary limitations. 
First, it was limited to two real collaborative work sessions in the field of pneumatic tire design. 
Second, although we have presented quantitative evidence, user feedback was essentially subjective. 
The next step is to perform a longitudinal research to measure the learnability and user adoption of WoW.
We also want to investigate the benefits of persisting meeting documents across consecutive sessions (\hyperref[par:time_config]{D5}).

Finally, we envision further developments of WoW by leveraging our findings (Section \ref{sec:findings}) to improve interface design, interaction controls, and better support for mixed-presence meetings. 
We are also currently extending WoW to support assembling coordinated multiple view visualization setups. 
WoW will evolve to support the integration of a broader range of data visualization components, thus allowing its use in other application domains.


\bibliographystyle{ACM-Reference-Format}
\bibliography{bibliography}


\begin{thebibliography}{83}


\ifx \showCODEN    \undefined \def \showCODEN     #1{\unskip}     \fi
\ifx \showDOI      \undefined \def \showDOI       #1{#1}\fi
\ifx \showISBNx    \undefined \def \showISBNx     #1{\unskip}     \fi
\ifx \showISBNxiii \undefined \def \showISBNxiii  #1{\unskip}     \fi
\ifx \showISSN     \undefined \def \showISSN      #1{\unskip}     \fi
\ifx \showLCCN     \undefined \def \showLCCN      #1{\unskip}     \fi
\ifx \shownote     \undefined \def \shownote      #1{#1}          \fi
\ifx \showarticletitle \undefined \def \showarticletitle #1{#1}   \fi
\ifx \showURL      \undefined \def \showURL       {\relax}        \fi
\providecommand\bibfield[2]{#2}
\providecommand\bibinfo[2]{#2}
\providecommand\natexlab[1]{#1}
\providecommand\showeprint[2][]{arXiv:#2}

\bibitem[\protect\citeauthoryear{Avellino, Fleury, Mackay, and
  Beaudouin-Lafon}{Avellino et~al\mbox{.}}{2017}]%
        {avellino2017camray}
\bibfield{author}{\bibinfo{person}{Ignacio Avellino},
  \bibinfo{person}{C{\'e}dric Fleury}, \bibinfo{person}{Wendy~E Mackay}, {and}
  \bibinfo{person}{Michel Beaudouin-Lafon}.} \bibinfo{year}{2017}\natexlab{}.
\newblock \showarticletitle{Camray: Camera arrays support remote collaboration
  on wall-sized displays}. In \bibinfo{booktitle}{\emph{Proceedings of the CHI
  Conference on Human Factors in Computing Systems}}. ACM,
  \bibinfo{pages}{6718--6729}.
\newblock
\urldef\tempurl%
\url{https://doi.org/10.1145/3025453.3025604}
\showDOI{\tempurl}


\bibitem[\protect\citeauthoryear{Badam, Amini, Elmqvist, and Irani}{Badam
  et~al\mbox{.}}{2016}]%
        {badam2016supporting}
\bibfield{author}{\bibinfo{person}{Sriram~Karthik Badam},
  \bibinfo{person}{Fereshteh Amini}, \bibinfo{person}{Niklas Elmqvist}, {and}
  \bibinfo{person}{Pourang Irani}.} \bibinfo{year}{2016}\natexlab{}.
\newblock \showarticletitle{Supporting visual exploration for multiple users in
  large display environments}. In \bibinfo{booktitle}{\emph{2016 IEEE
  Conference on Visual Analytics Science and Technology}}. IEEE,
  \bibinfo{pages}{1--10}.
\newblock
\urldef\tempurl%
\url{https://doi.org/10.1109/VAST.2016.7883506}
\showDOI{\tempurl}


\bibitem[\protect\citeauthoryear{Bafoutsou and Mentzas}{Bafoutsou and
  Mentzas}{2001}]%
        {bafoutsou2001comparative}
\bibfield{author}{\bibinfo{person}{Georgia Bafoutsou} {and}
  \bibinfo{person}{Gregoris Mentzas}.} \bibinfo{year}{2001}\natexlab{}.
\newblock \showarticletitle{A comparative analysis of web-based collaborative
  systems}. In \bibinfo{booktitle}{\emph{12th International Workshop on
  Database and Expert Systems Applications}}. IEEE, \bibinfo{pages}{496--500}.
\newblock


\bibitem[\protect\citeauthoryear{Ball, North, and Bowman}{Ball
  et~al\mbox{.}}{2007}]%
        {ball2007move}
\bibfield{author}{\bibinfo{person}{Robert Ball}, \bibinfo{person}{Chris North},
  {and} \bibinfo{person}{Doug~A Bowman}.} \bibinfo{year}{2007}\natexlab{}.
\newblock \showarticletitle{Move to improve: promoting physical navigation to
  increase user performance with large displays}. In
  \bibinfo{booktitle}{\emph{Proceedings of the SIGCHI conference on Human
  factors in computing systems}}. ACM, \bibinfo{pages}{191--200}.
\newblock
\urldef\tempurl%
\url{https://doi.org/10.1145/1240624.1240656}
\showDOI{\tempurl}


\bibitem[\protect\citeauthoryear{Bangor, Kortum, and Miller}{Bangor
  et~al\mbox{.}}{2008}]%
        {bangor2008empirical}
\bibfield{author}{\bibinfo{person}{Aaron Bangor}, \bibinfo{person}{Philip~T
  Kortum}, {and} \bibinfo{person}{James~T Miller}.}
  \bibinfo{year}{2008}\natexlab{}.
\newblock \showarticletitle{An empirical evaluation of the system usability
  scale}.
\newblock \bibinfo{journal}{\emph{Intl. Journal of Human--Computer
  Interaction}} \bibinfo{volume}{24}, \bibinfo{number}{6}
  (\bibinfo{year}{2008}), \bibinfo{pages}{574--594}.
\newblock


\bibitem[\protect\citeauthoryear{Bardram, Gueddana, Houben, and
  Nielsen}{Bardram et~al\mbox{.}}{2012}]%
        {bardram2012reticularspaces}
\bibfield{author}{\bibinfo{person}{Jakob Bardram}, \bibinfo{person}{Sofiane
  Gueddana}, \bibinfo{person}{Steven Houben}, {and} \bibinfo{person}{S{\o}ren
  Nielsen}.} \bibinfo{year}{2012}\natexlab{}.
\newblock \showarticletitle{ReticularSpaces: activity-based computing support
  for physically distributed and collaborative smart spaces}. In
  \bibinfo{booktitle}{\emph{Proceedings of the SIGCHI Conference on Human
  Factors in Computing Systems}}. \bibinfo{pages}{2845--2854}.
\newblock


\bibitem[\protect\citeauthoryear{Beaudouin-Lafon, Huot, Nancel, Mackay,
  Pietriga, Primet, Wagner, Chapuis, Pillias, Eagan,
  et~al\mbox{.}}{Beaudouin-Lafon et~al\mbox{.}}{2012}]%
        {beaudouin2012multisurface}
\bibfield{author}{\bibinfo{person}{Michel Beaudouin-Lafon},
  \bibinfo{person}{Stephane Huot}, \bibinfo{person}{Mathieu Nancel},
  \bibinfo{person}{Wendy Mackay}, \bibinfo{person}{Emmanuel Pietriga},
  \bibinfo{person}{Romain Primet}, \bibinfo{person}{Julie Wagner},
  \bibinfo{person}{Olivier Chapuis}, \bibinfo{person}{Clement Pillias},
  \bibinfo{person}{James Eagan}, {et~al\mbox{.}}}
  \bibinfo{year}{2012}\natexlab{}.
\newblock \showarticletitle{Multisurface interaction in the {WILD} room}.
\newblock \bibinfo{journal}{\emph{Computer}} \bibinfo{volume}{45},
  \bibinfo{number}{4} (\bibinfo{year}{2012}), \bibinfo{pages}{48--56}.
\newblock
\urldef\tempurl%
\url{https://doi.org/10.1109/MC.2012.110}
\showDOI{\tempurl}


\bibitem[\protect\citeauthoryear{Benishek and Lazzara}{Benishek and
  Lazzara}{2019}]%
        {benishek2019teams}
\bibfield{author}{\bibinfo{person}{Lauren~E Benishek} {and}
  \bibinfo{person}{Elizabeth~H Lazzara}.} \bibinfo{year}{2019}\natexlab{}.
\newblock \showarticletitle{Teams in a new era: Some considerations and
  implications}.
\newblock \bibinfo{journal}{\emph{Frontiers in psychology}}
  \bibinfo{volume}{10} (\bibinfo{year}{2019}), \bibinfo{pages}{1006}.
\newblock


\bibitem[\protect\citeauthoryear{Bi and Balakrishnan}{Bi and
  Balakrishnan}{2009}]%
        {bi2009comparing}
\bibfield{author}{\bibinfo{person}{Xiaojun Bi} {and} \bibinfo{person}{Ravin
  Balakrishnan}.} \bibinfo{year}{2009}\natexlab{}.
\newblock \showarticletitle{Comparing usage of a large high-resolution display
  to single or dual desktop displays for daily work}. In
  \bibinfo{booktitle}{\emph{Proceedings of the SIGCHI Conference on Human
  Factors in Computing Systems}}. ACM, \bibinfo{pages}{1005--1014}.
\newblock
\urldef\tempurl%
\url{https://doi.org/10.1145/1518701.1518855}
\showDOI{\tempurl}


\bibitem[\protect\citeauthoryear{Bragdon, DeLine, Hinckley, and Morris}{Bragdon
  et~al\mbox{.}}{2011}]%
        {bragdon2011code}
\bibfield{author}{\bibinfo{person}{Andrew Bragdon}, \bibinfo{person}{Rob
  DeLine}, \bibinfo{person}{Ken Hinckley}, {and}
  \bibinfo{person}{Meredith~Ringel Morris}.} \bibinfo{year}{2011}\natexlab{}.
\newblock \showarticletitle{Code space: touch+ air gesture hybrid interactions
  for supporting developer meetings}. In \bibinfo{booktitle}{\emph{Proceedings
  of the ACM International Conference on Interactive Tabletops and Surfaces}}.
  \bibinfo{pages}{212--221}.
\newblock


\bibitem[\protect\citeauthoryear{Brooke}{Brooke}{2013}]%
        {brooke2013sus}
\bibfield{author}{\bibinfo{person}{John Brooke}.}
  \bibinfo{year}{2013}\natexlab{}.
\newblock \showarticletitle{SUS: a retrospective}.
\newblock \bibinfo{journal}{\emph{Journal of usability studies}}
  \bibinfo{volume}{8}, \bibinfo{number}{2} (\bibinfo{year}{2013}),
  \bibinfo{pages}{29--40}.
\newblock


\bibitem[\protect\citeauthoryear{Carmeli, Gelbard, and Reiter-Palmon}{Carmeli
  et~al\mbox{.}}{2013}]%
        {carmeli2013leadership}
\bibfield{author}{\bibinfo{person}{Abraham Carmeli}, \bibinfo{person}{Roy
  Gelbard}, {and} \bibinfo{person}{Roni Reiter-Palmon}.}
  \bibinfo{year}{2013}\natexlab{}.
\newblock \showarticletitle{Leadership, creative problem-solving capacity, and
  creative performance: The importance of knowledge sharing}.
\newblock \bibinfo{journal}{\emph{Human Resource Management}}
  \bibinfo{volume}{52}, \bibinfo{number}{1} (\bibinfo{year}{2013}),
  \bibinfo{pages}{95--121}.
\newblock


\bibitem[\protect\citeauthoryear{Chan, Anslow, Seyed, and Maurer}{Chan
  et~al\mbox{.}}{2016}]%
        {chan2016envisioning}
\bibfield{author}{\bibinfo{person}{Edwin Chan}, \bibinfo{person}{Craig Anslow},
  \bibinfo{person}{Teddy Seyed}, {and} \bibinfo{person}{Frank Maurer}.}
  \bibinfo{year}{2016}\natexlab{}.
\newblock \showarticletitle{Envisioning the Emergency Operations Centre of the
  Future}.
\newblock In \bibinfo{booktitle}{\emph{Collaboration Meets Interactive
  Spaces}}. \bibinfo{publisher}{Springer}, \bibinfo{pages}{349--372}.
\newblock
\urldef\tempurl%
\url{https://doi.org/10.1007/978-3-319-45853-3_15}
\showDOI{\tempurl}


\bibitem[\protect\citeauthoryear{Chattopadhyay}{Chattopadhyay}{2018}]%
        {chattopadhyay2018shared}
\bibfield{author}{\bibinfo{person}{Debaleena Chattopadhyay}.}
  \bibinfo{year}{2018}\natexlab{}.
\newblock \showarticletitle{Shared Document Control in Multi-Device
  Classrooms}.
\newblock \bibinfo{journal}{\emph{Technical Report CI-MDC-10-2018, University
  of Illinois}} (\bibinfo{year}{2018}).
\newblock
\urldef\tempurl%
\url{https://doi.org/10.13140/RG.2.2.18321.68962}
\showDOI{\tempurl}


\bibitem[\protect\citeauthoryear{Chokshi, Seyed, Marinho~Rodrigues, and
  Maurer}{Chokshi et~al\mbox{.}}{2014}]%
        {chokshi2014eplan}
\bibfield{author}{\bibinfo{person}{Apoorve Chokshi}, \bibinfo{person}{Teddy
  Seyed}, \bibinfo{person}{Francisco Marinho~Rodrigues}, {and}
  \bibinfo{person}{Frank Maurer}.} \bibinfo{year}{2014}\natexlab{}.
\newblock \showarticletitle{ePlan multi-surface: A multi-surface environment
  for emergency response planning exercises}. In
  \bibinfo{booktitle}{\emph{Proceedings of the Ninth ACM International
  Conference on Interactive Tabletops and Surfaces}}. ACM,
  \bibinfo{pages}{219--228}.
\newblock
\urldef\tempurl%
\url{https://doi.org/10.1145/2669485.2669520}
\showDOI{\tempurl}


\bibitem[\protect\citeauthoryear{Clayphan, Martinez-Maldonado, Tomitsch,
  Atkinson, and Kay}{Clayphan et~al\mbox{.}}{2016}]%
        {clayphan2016wild}
\bibfield{author}{\bibinfo{person}{Andrew Clayphan}, \bibinfo{person}{Roberto
  Martinez-Maldonado}, \bibinfo{person}{Martin Tomitsch},
  \bibinfo{person}{Susan Atkinson}, {and} \bibinfo{person}{Judy Kay}.}
  \bibinfo{year}{2016}\natexlab{}.
\newblock \showarticletitle{An in-the-wild study of learning to brainstorm:
  comparing cards, tabletops and wall displays in the classroom}.
\newblock \bibinfo{journal}{\emph{Interacting with Computers}}
  \bibinfo{volume}{28}, \bibinfo{number}{6} (\bibinfo{year}{2016}),
  \bibinfo{pages}{788--810}.
\newblock
\urldef\tempurl%
\url{https://doi.org/10.1093/iwc/iww001}
\showDOI{\tempurl}


\bibitem[\protect\citeauthoryear{Cohen and Levesque}{Cohen and
  Levesque}{1991}]%
        {cohen1991teamwork}
\bibfield{author}{\bibinfo{person}{Philip~R Cohen} {and}
  \bibinfo{person}{Hector~J Levesque}.} \bibinfo{year}{1991}\natexlab{}.
\newblock \showarticletitle{Teamwork}.
\newblock \bibinfo{journal}{\emph{Nous}} \bibinfo{volume}{25},
  \bibinfo{number}{4} (\bibinfo{year}{1991}), \bibinfo{pages}{487--512}.
\newblock


\bibitem[\protect\citeauthoryear{Conley, Foley, Gorman, Denham, and
  Coleman}{Conley et~al\mbox{.}}{2017}]%
        {conley2017acquisition}
\bibfield{author}{\bibinfo{person}{Shannon~Nicole Conley},
  \bibinfo{person}{Rider~W Foley}, \bibinfo{person}{Michael~E Gorman},
  \bibinfo{person}{Jessica Denham}, {and} \bibinfo{person}{Kevin Coleman}.}
  \bibinfo{year}{2017}\natexlab{}.
\newblock \showarticletitle{Acquisition of T-shaped expertise: an exploratory
  study}.
\newblock \bibinfo{journal}{\emph{Social Epistemology}} \bibinfo{volume}{31},
  \bibinfo{number}{2} (\bibinfo{year}{2017}), \bibinfo{pages}{165--183}.
\newblock


\bibitem[\protect\citeauthoryear{Davidson}{Davidson}{2003}]%
        {davidson2003insights}
\bibfield{author}{\bibinfo{person}{Janet~E Davidson}.}
  \bibinfo{year}{2003}\natexlab{}.
\newblock \showarticletitle{Insights about insightful problem solving}.
\newblock \bibinfo{journal}{\emph{The psychology of problem solving}}
  (\bibinfo{year}{2003}), \bibinfo{pages}{149--175}.
\newblock


\bibitem[\protect\citeauthoryear{DiMicco, Pandolfo, and Bender}{DiMicco
  et~al\mbox{.}}{2004}]%
        {dimicco2004influencing}
\bibfield{author}{\bibinfo{person}{Joan~Morris DiMicco}, \bibinfo{person}{Anna
  Pandolfo}, {and} \bibinfo{person}{Walter Bender}.}
  \bibinfo{year}{2004}\natexlab{}.
\newblock \showarticletitle{Influencing group participation with a shared
  display}. In \bibinfo{booktitle}{\emph{Proceedings of the 2004 ACM conference
  on Computer supported cooperative work}}. \bibinfo{pages}{614--623}.
\newblock


\bibitem[\protect\citeauthoryear{Doshi, Tuteja, Bharadwaj, Tantillo, Marrinan,
  Patton, and Marai}{Doshi et~al\mbox{.}}{2017}]%
        {doshi2017stickyschedule}
\bibfield{author}{\bibinfo{person}{Vishal Doshi}, \bibinfo{person}{Sneha
  Tuteja}, \bibinfo{person}{Krishna Bharadwaj}, \bibinfo{person}{Davide
  Tantillo}, \bibinfo{person}{Thomas Marrinan}, \bibinfo{person}{James Patton},
  {and} \bibinfo{person}{G~Elisabeta Marai}.} \bibinfo{year}{2017}\natexlab{}.
\newblock \showarticletitle{StickySchedule: an interactive multi-user
  application for conference scheduling on large-scale shared displays}. In
  \bibinfo{booktitle}{\emph{Proceedings of the 6th ACM International Symposium
  on Pervasive Displays}}. ACM, \bibinfo{pages}{2}.
\newblock
\urldef\tempurl%
\url{https://doi.org/10.1145/3078810.3078817}
\showDOI{\tempurl}


\bibitem[\protect\citeauthoryear{Driskell, Salas, and Driskell}{Driskell
  et~al\mbox{.}}{2018}]%
        {driskell2018foundations}
\bibfield{author}{\bibinfo{person}{James~E Driskell}, \bibinfo{person}{Eduardo
  Salas}, {and} \bibinfo{person}{Tripp Driskell}.}
  \bibinfo{year}{2018}\natexlab{}.
\newblock \showarticletitle{Foundations of teamwork and collaboration.}
\newblock \bibinfo{journal}{\emph{American Psychologist}} \bibinfo{volume}{73},
  \bibinfo{number}{4} (\bibinfo{year}{2018}), \bibinfo{pages}{334}.
\newblock


\bibitem[\protect\citeauthoryear{Erlingsson and Brysiewicz}{Erlingsson and
  Brysiewicz}{2017}]%
        {erlingsson2017hands}
\bibfield{author}{\bibinfo{person}{Christen Erlingsson} {and}
  \bibinfo{person}{Petra Brysiewicz}.} \bibinfo{year}{2017}\natexlab{}.
\newblock \showarticletitle{A hands-on guide to doing content analysis}.
\newblock \bibinfo{journal}{\emph{African Journal of Emergency Medicine}}
  \bibinfo{volume}{7}, \bibinfo{number}{3} (\bibinfo{year}{2017}),
  \bibinfo{pages}{93--99}.
\newblock


\bibitem[\protect\citeauthoryear{Feibush, Gagvani, and Williams}{Feibush
  et~al\mbox{.}}{2000}]%
        {feibush2000visualization}
\bibfield{author}{\bibinfo{person}{Eliot Feibush}, \bibinfo{person}{Nikhil
  Gagvani}, {and} \bibinfo{person}{Daniel Williams}.}
  \bibinfo{year}{2000}\natexlab{}.
\newblock \showarticletitle{Visualization for situational awareness}.
\newblock \bibinfo{journal}{\emph{IEEE Computer Graphics and Applications}}
  \bibinfo{volume}{20}, \bibinfo{number}{5} (\bibinfo{year}{2000}),
  \bibinfo{pages}{38--45}.
\newblock


\bibitem[\protect\citeauthoryear{Gardner}{Gardner}{2005}]%
        {gardner2005veritable}
\bibfield{author}{\bibinfo{person}{William Gardner}.}
  \bibinfo{year}{2005}\natexlab{}.
\newblock \showarticletitle{Veritable authentic leadership: Emergence,
  functioning and impacts}.
\newblock In \bibinfo{booktitle}{\emph{Veritable authentic leadership:
  Emergence, functioning and impacts}}. \bibinfo{publisher}{Elsevier},
  \bibinfo{pages}{3--41}.
\newblock


\bibitem[\protect\citeauthoryear{Gent and Walter}{Gent and Walter}{2006}]%
        {Gent2006}
\bibfield{author}{\bibinfo{person}{Alan~N Gent} {and} \bibinfo{person}{Joseph~D
  Walter}.} \bibinfo{year}{2006}\natexlab{}.
\newblock \bibinfo{booktitle}{\emph{{The Pneumatic Tyre}}}.
\newblock Number February.
\newblock


\bibitem[\protect\citeauthoryear{Gutwin and Greenberg}{Gutwin and
  Greenberg}{2004}]%
        {gutwin2004importance}
\bibfield{author}{\bibinfo{person}{Carl Gutwin} {and} \bibinfo{person}{Saul
  Greenberg}.} \bibinfo{year}{2004}\natexlab{}.
\newblock \showarticletitle{The importance of awareness for team cognition in
  distributed collaboration.}
\newblock  (\bibinfo{year}{2004}).
\newblock


\bibitem[\protect\citeauthoryear{Hart}{Hart}{2006}]%
        {hart2006nasa}
\bibfield{author}{\bibinfo{person}{Sandra~G Hart}.}
  \bibinfo{year}{2006}\natexlab{}.
\newblock \showarticletitle{NASA-task load index (NASA-TLX); 20 years later}.
  In \bibinfo{booktitle}{\emph{Proceedings of the human factors and ergonomics
  society annual meeting}}, Vol.~\bibinfo{volume}{50}. Sage publications Sage
  CA: Los Angeles, CA, \bibinfo{pages}{904--908}.
\newblock


\bibitem[\protect\citeauthoryear{Hassan, Mahsud, Yukl, and Prussia}{Hassan
  et~al\mbox{.}}{2013}]%
        {hassan2013ethical}
\bibfield{author}{\bibinfo{person}{Shahidul Hassan},
  \bibinfo{person}{Rubin{\'a} Mahsud}, \bibinfo{person}{Gary Yukl}, {and}
  \bibinfo{person}{Gregory~E Prussia}.} \bibinfo{year}{2013}\natexlab{}.
\newblock \showarticletitle{Ethical and empowering leadership and leader
  effectiveness}.
\newblock \bibinfo{journal}{\emph{Journal of Managerial Psychology}}
  (\bibinfo{year}{2013}).
\newblock


\bibitem[\protect\citeauthoryear{He, Du, and Perlin}{He et~al\mbox{.}}{2020}]%
        {he2020collabovr}
\bibfield{author}{\bibinfo{person}{Zhenyi He}, \bibinfo{person}{Ruofei Du},
  {and} \bibinfo{person}{Ken Perlin}.} \bibinfo{year}{2020}\natexlab{}.
\newblock \showarticletitle{CollaboVR: A Reconfigurable Framework for Creative
  Collaboration in Virtual Reality}. In \bibinfo{booktitle}{\emph{2020 IEEE
  International Symposium on Mixed and Augmented Reality (ISMAR)}}. IEEE,
  \bibinfo{pages}{542--554}.
\newblock


\bibitem[\protect\citeauthoryear{Holyoak}{Holyoak}{1990}]%
        {holyoak1990problem}
\bibfield{author}{\bibinfo{person}{Keith~J Holyoak}.}
  \bibinfo{year}{1990}\natexlab{}.
\newblock \showarticletitle{Problem solving}.
\newblock \bibinfo{journal}{\emph{Thinking: An invitation to cognitive
  science}}  \bibinfo{volume}{3} (\bibinfo{year}{1990}),
  \bibinfo{pages}{117--146}.
\newblock


\bibitem[\protect\citeauthoryear{Hughes, Lee, Tian, Newman, and Legood}{Hughes
  et~al\mbox{.}}{2018}]%
        {hughes2018leadership}
\bibfield{author}{\bibinfo{person}{David~J Hughes}, \bibinfo{person}{Allan
  Lee}, \bibinfo{person}{Amy~Wei Tian}, \bibinfo{person}{Alex Newman}, {and}
  \bibinfo{person}{Alison Legood}.} \bibinfo{year}{2018}\natexlab{}.
\newblock \showarticletitle{Leadership, creativity, and innovation: A critical
  review and practical recommendations}.
\newblock \bibinfo{journal}{\emph{The Leadership Quarterly}}
  \bibinfo{volume}{29}, \bibinfo{number}{5} (\bibinfo{year}{2018}),
  \bibinfo{pages}{549--569}.
\newblock


\bibitem[\protect\citeauthoryear{Hutchins}{Hutchins}{1995}]%
        {hutchins1995cognition}
\bibfield{author}{\bibinfo{person}{Edwin Hutchins}.}
  \bibinfo{year}{1995}\natexlab{}.
\newblock \bibinfo{booktitle}{\emph{Cognition in the Wild}}.
\newblock Number 1995. \bibinfo{publisher}{MIT press}.
\newblock


\bibitem[\protect\citeauthoryear{Ion, Chang, Haller, Hancock, and Scott}{Ion
  et~al\mbox{.}}{2013}]%
        {ion2013canyon}
\bibfield{author}{\bibinfo{person}{Alexandra Ion}, \bibinfo{person}{Y-L~Betty
  Chang}, \bibinfo{person}{Michael Haller}, \bibinfo{person}{Mark Hancock},
  {and} \bibinfo{person}{Stacey~D Scott}.} \bibinfo{year}{2013}\natexlab{}.
\newblock \showarticletitle{Canyon: providing location awareness of multiple
  moving objects in a detail view on large displays}. In
  \bibinfo{booktitle}{\emph{Proceedings of the SIGCHI Conference on Human
  Factors in Computing Systems}}. ACM, \bibinfo{pages}{3149--3158}.
\newblock
\urldef\tempurl%
\url{https://doi.org/10.1145/2470654.2466431}
\showDOI{\tempurl}


\bibitem[\protect\citeauthoryear{Isenberg, Fisher, Paul, Morris, Inkpen, and
  Czerwinski}{Isenberg et~al\mbox{.}}{2011}]%
        {isenberg2011co}
\bibfield{author}{\bibinfo{person}{Petra Isenberg}, \bibinfo{person}{Danyel
  Fisher}, \bibinfo{person}{Sharoda~A Paul}, \bibinfo{person}{Meredith~Ringel
  Morris}, \bibinfo{person}{Kori Inkpen}, {and} \bibinfo{person}{Mary
  Czerwinski}.} \bibinfo{year}{2011}\natexlab{}.
\newblock \showarticletitle{Co-located collaborative visual analytics around a
  tabletop display}.
\newblock \bibinfo{journal}{\emph{IEEE Transactions on visualization and
  Computer Graphics}} \bibinfo{volume}{18}, \bibinfo{number}{5}
  (\bibinfo{year}{2011}), \bibinfo{pages}{689--702}.
\newblock


\bibitem[\protect\citeauthoryear{Jakobsen, Haile, Knudsen, and
  Hornb{\ae}k}{Jakobsen et~al\mbox{.}}{2013}]%
        {jakobsen2013information}
\bibfield{author}{\bibinfo{person}{Mikkel~R Jakobsen},
  \bibinfo{person}{Yonas~Sahlemariam Haile}, \bibinfo{person}{S{\o}ren
  Knudsen}, {and} \bibinfo{person}{Kasper Hornb{\ae}k}.}
  \bibinfo{year}{2013}\natexlab{}.
\newblock \showarticletitle{Information visualization and proxemics: Design
  opportunities and empirical findings}.
\newblock \bibinfo{journal}{\emph{IEEE Transactions on Visualization and
  Computer Graphics}} \bibinfo{volume}{19}, \bibinfo{number}{12}
  (\bibinfo{year}{2013}), \bibinfo{pages}{2386--2395}.
\newblock
\urldef\tempurl%
\url{https://doi.org/10.1109/TVCG.2013.166}
\showDOI{\tempurl}


\bibitem[\protect\citeauthoryear{Jakobsen and Hornb{\ae}k}{Jakobsen and
  Hornb{\ae}k}{2014}]%
        {jakobsen2014up}
\bibfield{author}{\bibinfo{person}{Mikkel~R Jakobsen} {and}
  \bibinfo{person}{Kasper Hornb{\ae}k}.} \bibinfo{year}{2014}\natexlab{}.
\newblock \showarticletitle{Up close and personal: Collaborative work on a
  high-resolution multitouch wall display}.
\newblock \bibinfo{journal}{\emph{ACM Transactions on Computer-Human
  Interaction (TOCHI)}} \bibinfo{volume}{21}, \bibinfo{number}{2}
  (\bibinfo{year}{2014}), \bibinfo{pages}{11}.
\newblock
\urldef\tempurl%
\url{https://doi.org/10.1145/2576099}
\showDOI{\tempurl}


\bibitem[\protect\citeauthoryear{Jakobsen and Hornb{\ae}k}{Jakobsen and
  Hornb{\ae}k}{2016}]%
        {jakobsen2016negotiating}
\bibfield{author}{\bibinfo{person}{Mikkel~R. Jakobsen} {and}
  \bibinfo{person}{Kasper Hornb{\ae}k}.} \bibinfo{year}{2016}\natexlab{}.
\newblock \showarticletitle{Negotiating for Space?: Collaborative Work Using a
  Wall Display with Mouse and Touch Input}. In
  \bibinfo{booktitle}{\emph{Proceedings of the CHI Conference on Human Factors
  in Computing Systems}} (San Jose, California, USA)
  \emph{(\bibinfo{series}{CHI '16})}. \bibinfo{publisher}{ACM},
  \bibinfo{pages}{2050--2061}.
\newblock
\showISBNx{978-1-4503-3362-7}
\urldef\tempurl%
\url{https://doi.org/10.1145/2858036.2858158}
\showDOI{\tempurl}


\bibitem[\protect\citeauthoryear{Jansen, Schjerlund, and Hornb{\ae}k}{Jansen
  et~al\mbox{.}}{2019}]%
        {jansen2019effects}
\bibfield{author}{\bibinfo{person}{Yvonne Jansen}, \bibinfo{person}{Jonas
  Schjerlund}, {and} \bibinfo{person}{Kasper Hornb{\ae}k}.}
  \bibinfo{year}{2019}\natexlab{}.
\newblock \showarticletitle{Effects of Locomotion and Visual Overview on
  Spatial Memory when Interacting with Wall Displays}. In
  \bibinfo{booktitle}{\emph{Proceedings of the CHI Conference on Human Factors
  in Computing Systems}}. ACM, \bibinfo{pages}{291}.
\newblock
\urldef\tempurl%
\url{https://doi.org/10.1145/3290605.3300521}
\showDOI{\tempurl}


\bibitem[\protect\citeauthoryear{Khan, Matejka, Fitzmaurice, Kurtenbach,
  Burtnyk, and Buxton}{Khan et~al\mbox{.}}{2009}]%
        {khan2009toward}
\bibfield{author}{\bibinfo{person}{Azam Khan}, \bibinfo{person}{Justin
  Matejka}, \bibinfo{person}{George Fitzmaurice}, \bibinfo{person}{Gord
  Kurtenbach}, \bibinfo{person}{Nicolas Burtnyk}, {and} \bibinfo{person}{Bill
  Buxton}.} \bibinfo{year}{2009}\natexlab{}.
\newblock \showarticletitle{Toward the digital design studio: Large display
  explorations}.
\newblock \bibinfo{journal}{\emph{Human--Computer Interaction}}
  \bibinfo{volume}{24}, \bibinfo{number}{1-2} (\bibinfo{year}{2009}),
  \bibinfo{pages}{9--47}.
\newblock
\urldef\tempurl%
\url{https://doi.org/10.1080/07370020902819932}
\showDOI{\tempurl}


\bibitem[\protect\citeauthoryear{Kister, Klamka, Tominski, and Dachselt}{Kister
  et~al\mbox{.}}{2017}]%
        {kister2017grasp}
\bibfield{author}{\bibinfo{person}{Ulrike Kister}, \bibinfo{person}{Konstantin
  Klamka}, \bibinfo{person}{Christian Tominski}, {and} \bibinfo{person}{Raimund
  Dachselt}.} \bibinfo{year}{2017}\natexlab{}.
\newblock \showarticletitle{GraSp: Combining Spatially-aware Mobile Devices and
  a Display Wall for Graph Visualization and Interaction}.
\newblock \bibinfo{journal}{\emph{Computer Graphics Forum}}
  \bibinfo{volume}{36}, \bibinfo{number}{3} (\bibinfo{year}{2017}),
  \bibinfo{pages}{503--514}.
\newblock
\urldef\tempurl%
\url{https://doi.org/10.1111/cgf.13206}
\showDOI{\tempurl}


\bibitem[\protect\citeauthoryear{Kister, Reipschl{\"a}ger, Matulic, and
  Dachselt}{Kister et~al\mbox{.}}{2015}]%
        {kister2015bodylenses}
\bibfield{author}{\bibinfo{person}{Ulrike Kister}, \bibinfo{person}{Patrick
  Reipschl{\"a}ger}, \bibinfo{person}{Fabrice Matulic}, {and}
  \bibinfo{person}{Raimund Dachselt}.} \bibinfo{year}{2015}\natexlab{}.
\newblock \showarticletitle{BodyLenses: Embodied magic lenses and personal
  territories for wall displays}. In \bibinfo{booktitle}{\emph{Proceedings of
  the International Conference on Interactive Tabletops \& Surfaces}}. ACM,
  \bibinfo{pages}{117--126}.
\newblock
\urldef\tempurl%
\url{https://doi.org/10.1145/2817721.2817726}
\showDOI{\tempurl}


\bibitem[\protect\citeauthoryear{Langner, Kister, and Dachselt}{Langner
  et~al\mbox{.}}{2018}]%
        {langner2018multiple}
\bibfield{author}{\bibinfo{person}{Ricardo Langner}, \bibinfo{person}{Ulrike
  Kister}, {and} \bibinfo{person}{Raimund Dachselt}.}
  \bibinfo{year}{2018}\natexlab{}.
\newblock \showarticletitle{Multiple coordinated views at large displays for
  multiple users: Empirical findings on user behavior, movements, and
  distances}.
\newblock \bibinfo{journal}{\emph{IEEE Transactions on Visualization and
  Computer Graphics}} \bibinfo{volume}{25}, \bibinfo{number}{1}
  (\bibinfo{year}{2018}), \bibinfo{pages}{608--618}.
\newblock
\urldef\tempurl%
\url{https://doi.org/10.1109/TVCG.2018.2865235}
\showDOI{\tempurl}


\bibitem[\protect\citeauthoryear{Leigh, Johnson, Renambot, Peterka, Jeong,
  Sandin, Talandis, Jagodic, Nam, Hur, et~al\mbox{.}}{Leigh
  et~al\mbox{.}}{2012}]%
        {leigh2012scalable}
\bibfield{author}{\bibinfo{person}{Jason Leigh}, \bibinfo{person}{Andrew
  Johnson}, \bibinfo{person}{Luc Renambot}, \bibinfo{person}{Tom Peterka},
  \bibinfo{person}{Byungil Jeong}, \bibinfo{person}{Daniel~J Sandin},
  \bibinfo{person}{Jonas Talandis}, \bibinfo{person}{Ratko Jagodic},
  \bibinfo{person}{Sungwon Nam}, \bibinfo{person}{Hyejung Hur},
  {et~al\mbox{.}}} \bibinfo{year}{2012}\natexlab{}.
\newblock \showarticletitle{Scalable resolution display walls}.
\newblock \bibinfo{journal}{\emph{Proc. IEEE}} \bibinfo{volume}{101},
  \bibinfo{number}{1} (\bibinfo{year}{2012}), \bibinfo{pages}{115--129}.
\newblock
\urldef\tempurl%
\url{https://doi.org/10.1109/JPROC.2012.2191609}
\showDOI{\tempurl}


\bibitem[\protect\citeauthoryear{Lischke, Gr{\"u}ninger, Klouche, Schmidt,
  Slusallek, and Jacucci}{Lischke et~al\mbox{.}}{2015}]%
        {lischke2015interaction}
\bibfield{author}{\bibinfo{person}{Lars Lischke}, \bibinfo{person}{J{\"u}rgen
  Gr{\"u}ninger}, \bibinfo{person}{Khalil Klouche}, \bibinfo{person}{Albrecht
  Schmidt}, \bibinfo{person}{Philipp Slusallek}, {and} \bibinfo{person}{Giulio
  Jacucci}.} \bibinfo{year}{2015}\natexlab{}.
\newblock \showarticletitle{Interaction techniques for wall-sized screens}. In
  \bibinfo{booktitle}{\emph{Proceedings of the International Conference on
  Interactive Tabletops \& Surfaces}}. ACM, \bibinfo{pages}{501--504}.
\newblock
\urldef\tempurl%
\url{https://doi.org/10.1145/2817721.2835071}
\showDOI{\tempurl}


\bibitem[\protect\citeauthoryear{Liu, Chapuis, Beaudouin-Lafon, and
  Lecolinet}{Liu et~al\mbox{.}}{2016}]%
        {liu2016shared}
\bibfield{author}{\bibinfo{person}{Can Liu}, \bibinfo{person}{Olivier Chapuis},
  \bibinfo{person}{Michel Beaudouin-Lafon}, {and} \bibinfo{person}{Eric
  Lecolinet}.} \bibinfo{year}{2016}\natexlab{}.
\newblock \showarticletitle{Shared interaction on a wall-sized display in a
  data manipulation task}. In \bibinfo{booktitle}{\emph{Proceedings of the CHI
  Conference on Human Factors in Computing Systems}}. ACM,
  \bibinfo{pages}{2075--2086}.
\newblock
\urldef\tempurl%
\url{https://doi.org/10.1145/2858036.2858039}
\showDOI{\tempurl}


\bibitem[\protect\citeauthoryear{Liu, Chapuis, Beaudouin-Lafon, and
  Lecolinet}{Liu et~al\mbox{.}}{2017}]%
        {liu2017coreach}
\bibfield{author}{\bibinfo{person}{Can Liu}, \bibinfo{person}{Olivier Chapuis},
  \bibinfo{person}{Michel Beaudouin-Lafon}, {and} \bibinfo{person}{Eric
  Lecolinet}.} \bibinfo{year}{2017}\natexlab{}.
\newblock \showarticletitle{CoReach: Cooperative gestures for data manipulation
  on wall-sized displays}. In \bibinfo{booktitle}{\emph{Proceedings of the CHI
  Conference on Human Factors in Computing Systems}}. ACM,
  \bibinfo{pages}{6730--6741}.
\newblock
\urldef\tempurl%
\url{https://doi.org/10.1145/3025453.3025594}
\showDOI{\tempurl}


\bibitem[\protect\citeauthoryear{Lu, Yuan, and McLeod}{Lu
  et~al\mbox{.}}{2012}]%
        {lu2012twenty}
\bibfield{author}{\bibinfo{person}{Li Lu}, \bibinfo{person}{Y~Connie Yuan},
  {and} \bibinfo{person}{Poppy~Lauretta McLeod}.}
  \bibinfo{year}{2012}\natexlab{}.
\newblock \showarticletitle{Twenty-five years of hidden profiles in group
  decision making: A meta-analysis}.
\newblock \bibinfo{journal}{\emph{Personality and Social Psychology Review}}
  \bibinfo{volume}{16}, \bibinfo{number}{1} (\bibinfo{year}{2012}),
  \bibinfo{pages}{54--75}.
\newblock


\bibitem[\protect\citeauthoryear{Mahyar, Burke, Xiang, Meng, Booth, Girling,
  and Kellett}{Mahyar et~al\mbox{.}}{2016}]%
        {mahyar2016ud}
\bibfield{author}{\bibinfo{person}{Narges Mahyar}, \bibinfo{person}{Kelly~J
  Burke}, \bibinfo{person}{Jialiang Xiang}, \bibinfo{person}{Siyi Meng},
  \bibinfo{person}{Kellogg~S Booth}, \bibinfo{person}{Cynthia~L Girling}, {and}
  \bibinfo{person}{Ronald~W Kellett}.} \bibinfo{year}{2016}\natexlab{}.
\newblock \showarticletitle{Ud co-spaces: A table-centred multi-display
  environment for public engagement in urban design charrettes}. In
  \bibinfo{booktitle}{\emph{Proceedings of the 2016 ACM international
  conference on interactive surfaces and spaces}}. \bibinfo{pages}{109--118}.
\newblock


\bibitem[\protect\citeauthoryear{Mangano, LaToza, Petre, and van~der
  Hoek}{Mangano et~al\mbox{.}}{2014}]%
        {mangano2014supporting}
\bibfield{author}{\bibinfo{person}{Nicolas Mangano}, \bibinfo{person}{Thomas~D
  LaToza}, \bibinfo{person}{Marian Petre}, {and} \bibinfo{person}{Andr{\'e}
  van~der Hoek}.} \bibinfo{year}{2014}\natexlab{}.
\newblock \showarticletitle{Supporting informal design with interactive
  whiteboards}. In \bibinfo{booktitle}{\emph{Proceedings of the SIGCHI
  Conference on Human Factors in Computing Systems}}.
  \bibinfo{pages}{331--340}.
\newblock


\bibitem[\protect\citeauthoryear{Marques-Quinteiro, Rico, Passos, and
  Curral}{Marques-Quinteiro et~al\mbox{.}}{2019}]%
        {marques2019there}
\bibfield{author}{\bibinfo{person}{Pedro Marques-Quinteiro},
  \bibinfo{person}{Ram{\'o}n Rico}, \bibinfo{person}{Ana~M Passos}, {and}
  \bibinfo{person}{Lu{\'\i}s Curral}.} \bibinfo{year}{2019}\natexlab{}.
\newblock \showarticletitle{There is light and there is darkness: On the
  temporal dynamics of cohesion, coordination, and performance in business
  teams}.
\newblock \bibinfo{journal}{\emph{Frontiers in psychology}}
  \bibinfo{volume}{10} (\bibinfo{year}{2019}), \bibinfo{pages}{847}.
\newblock


\bibitem[\protect\citeauthoryear{Marrinan, Aurisano, Nishimoto, Bharadwaj,
  Mateevitsi, Renambot, Long, Johnson, and Leigh}{Marrinan
  et~al\mbox{.}}{2014}]%
        {marrinan2014sage2}
\bibfield{author}{\bibinfo{person}{Thomas Marrinan}, \bibinfo{person}{Jillian
  Aurisano}, \bibinfo{person}{Arthur Nishimoto}, \bibinfo{person}{Krishna
  Bharadwaj}, \bibinfo{person}{Victor Mateevitsi}, \bibinfo{person}{Luc
  Renambot}, \bibinfo{person}{Lance Long}, \bibinfo{person}{Andrew Johnson},
  {and} \bibinfo{person}{Jason Leigh}.} \bibinfo{year}{2014}\natexlab{}.
\newblock \showarticletitle{SAGE2: A new approach for data intensive
  collaboration using Scalable Resolution Shared Displays}. In
  \bibinfo{booktitle}{\emph{10th IEEE International Conference on Collaborative
  Computing: Networking, Applications and Worksharing}}. IEEE,
  \bibinfo{pages}{177--186}.
\newblock


\bibitem[\protect\citeauthoryear{Mateescu, Pimmer, Zahn, Klinkhammer, and
  Reiterer}{Mateescu et~al\mbox{.}}{2021}]%
        {mateescu2021collaboration}
\bibfield{author}{\bibinfo{person}{Magdalena Mateescu},
  \bibinfo{person}{Christoph Pimmer}, \bibinfo{person}{Carmen Zahn},
  \bibinfo{person}{Daniel Klinkhammer}, {and} \bibinfo{person}{Harald
  Reiterer}.} \bibinfo{year}{2021}\natexlab{}.
\newblock \showarticletitle{Collaboration on large interactive displays: a
  systematic review}.
\newblock \bibinfo{journal}{\emph{Human--Computer Interaction}}
  \bibinfo{volume}{36}, \bibinfo{number}{3} (\bibinfo{year}{2021}),
  \bibinfo{pages}{243--277}.
\newblock
\urldef\tempurl%
\url{https://doi.org/10.1080/07370024.2019.1697697}
\showDOI{\tempurl}


\bibitem[\protect\citeauthoryear{Mayer, Lischke, Gr{\o}nb{\ae}k, Sarsenbayeva,
  Vogelsang, Wo{\'z}niak, Henze, and Jacucci}{Mayer et~al\mbox{.}}{2018}]%
        {mayer2018pac}
\bibfield{author}{\bibinfo{person}{Sven Mayer}, \bibinfo{person}{Lars Lischke},
  \bibinfo{person}{Jens~Emil Gr{\o}nb{\ae}k}, \bibinfo{person}{Zhanna
  Sarsenbayeva}, \bibinfo{person}{Jonas Vogelsang}, \bibinfo{person}{Pawe{\l}~W
  Wo{\'z}niak}, \bibinfo{person}{Niels Henze}, {and} \bibinfo{person}{Giulio
  Jacucci}.} \bibinfo{year}{2018}\natexlab{}.
\newblock \showarticletitle{Pac-many: Movement behavior when playing
  collaborative and competitive games on large displays}. In
  \bibinfo{booktitle}{\emph{Proceedings of the CHI Conference on Human Factors
  in Computing Systems}}. ACM, \bibinfo{pages}{539}.
\newblock
\urldef\tempurl%
\url{https://doi.org/10.1145/3173574.3174113}
\showDOI{\tempurl}


\bibitem[\protect\citeauthoryear{Nguyen and Shanks}{Nguyen and Shanks}{2009}]%
        {nguyen2009framework}
\bibfield{author}{\bibinfo{person}{Lemai Nguyen} {and} \bibinfo{person}{Graeme
  Shanks}.} \bibinfo{year}{2009}\natexlab{}.
\newblock \showarticletitle{A framework for understanding creativity in
  requirements engineering}.
\newblock \bibinfo{journal}{\emph{Information and software technology}}
  \bibinfo{volume}{51}, \bibinfo{number}{3} (\bibinfo{year}{2009}),
  \bibinfo{pages}{655--662}.
\newblock


\bibitem[\protect\citeauthoryear{Nolte, Brown, Anslow, Wiechers, Polyvyanyy,
  and Herrmann}{Nolte et~al\mbox{.}}{2016}]%
        {nolte2016collaborative}
\bibfield{author}{\bibinfo{person}{Alexander Nolte}, \bibinfo{person}{Ross
  Brown}, \bibinfo{person}{Craig Anslow}, \bibinfo{person}{Moritz Wiechers},
  \bibinfo{person}{Artem Polyvyanyy}, {and} \bibinfo{person}{Thomas Herrmann}.}
  \bibinfo{year}{2016}\natexlab{}.
\newblock \showarticletitle{Collaborative business process modeling in
  multi-surface environments}.
\newblock In \bibinfo{booktitle}{\emph{Collaboration Meets Interactive
  Spaces}}. \bibinfo{publisher}{Springer}, \bibinfo{pages}{259--286}.
\newblock
\urldef\tempurl%
\url{https://doi.org/10.1007/978-3-319-45853-3_12}
\showDOI{\tempurl}


\bibitem[\protect\citeauthoryear{Novick and Bassok}{Novick and Bassok}{2005}]%
        {novick2005problem}
\bibfield{author}{\bibinfo{person}{Laura~R Novick} {and}
  \bibinfo{person}{Miriam Bassok}.} \bibinfo{year}{2005}\natexlab{}.
\newblock \bibinfo{booktitle}{\emph{Problem Solving.}}
\newblock \bibinfo{publisher}{Cambridge University Press}.
\newblock


\bibitem[\protect\citeauthoryear{Oliveri, Lawless, and Molloy}{Oliveri
  et~al\mbox{.}}{2017}]%
        {oliveri2017literature}
\bibfield{author}{\bibinfo{person}{Mar{\'\i}a~Elena Oliveri},
  \bibinfo{person}{Ren{\'e} Lawless}, {and} \bibinfo{person}{Hillary Molloy}.}
  \bibinfo{year}{2017}\natexlab{}.
\newblock \showarticletitle{A literature review on collaborative problem
  solving for college and workforce readiness}.
\newblock \bibinfo{journal}{\emph{ETS Research Report Series}}
  \bibinfo{volume}{2017}, \bibinfo{number}{1} (\bibinfo{year}{2017}),
  \bibinfo{pages}{1--27}.
\newblock


\bibitem[\protect\citeauthoryear{Onorati, Isenberg, Bezerianos, Pietriga, and
  Diaz}{Onorati et~al\mbox{.}}{2015}]%
        {onorati2015walltweet}
\bibfield{author}{\bibinfo{person}{Teresa Onorati}, \bibinfo{person}{Petra
  Isenberg}, \bibinfo{person}{Anastasia Bezerianos}, \bibinfo{person}{Emmanuel
  Pietriga}, {and} \bibinfo{person}{Paloma Diaz}.}
  \bibinfo{year}{2015}\natexlab{}.
\newblock \showarticletitle{WallTweet: A Knowledge Ecosystem for Supporting
  Situation Awareness}.
\newblock \bibinfo{journal}{\emph{Workshop on Data Exploration for Interactive
  Surfaces}} (\bibinfo{year}{2015}).
\newblock


\bibitem[\protect\citeauthoryear{Owens, Teves, Nguyen, Smith, Phelps, and
  Chaparro}{Owens et~al\mbox{.}}{2012}]%
        {owens2012examination}
\bibfield{author}{\bibinfo{person}{Justin~W Owens}, \bibinfo{person}{Jennifer
  Teves}, \bibinfo{person}{Bobby Nguyen}, \bibinfo{person}{Amanda Smith},
  \bibinfo{person}{Mandy~C Phelps}, {and} \bibinfo{person}{Barbara~S
  Chaparro}.} \bibinfo{year}{2012}\natexlab{}.
\newblock \showarticletitle{Examination of dual vs. single monitor use during
  common office tasks}. In \bibinfo{booktitle}{\emph{Proceedings of the Human
  Factors and Ergonomics Society Annual Meeting}}, Vol.~\bibinfo{volume}{56}.
  SAGE Publications Sage CA: Los Angeles, CA, \bibinfo{pages}{1506--1510}.
\newblock


\bibitem[\protect\citeauthoryear{Peltonen, Kurvinen, Salovaara, Jacucci,
  Ilmonen, Evans, Oulasvirta, and Saarikko}{Peltonen et~al\mbox{.}}{2008}]%
        {peltonen2008s}
\bibfield{author}{\bibinfo{person}{Peter Peltonen}, \bibinfo{person}{Esko
  Kurvinen}, \bibinfo{person}{Antti Salovaara}, \bibinfo{person}{Giulio
  Jacucci}, \bibinfo{person}{Tommi Ilmonen}, \bibinfo{person}{John Evans},
  \bibinfo{person}{Antti Oulasvirta}, {and} \bibinfo{person}{Petri Saarikko}.}
  \bibinfo{year}{2008}\natexlab{}.
\newblock \showarticletitle{It's Mine, Don't Touch!: interactions at a large
  multi-touch display in a city centre}. In
  \bibinfo{booktitle}{\emph{Proceedings of the SIGCHI conference on human
  factors in computing systems}}. ACM, \bibinfo{pages}{1285--1294}.
\newblock
\urldef\tempurl%
\url{https://doi.org/10.1145/1357054.1357255}
\showDOI{\tempurl}


\bibitem[\protect\citeauthoryear{Pietriga, Del~Campo, Ibsen, Primet, Appert,
  Chapuis, Hempel, Mu{\~n}oz, Eyheramendy, Jordan, et~al\mbox{.}}{Pietriga
  et~al\mbox{.}}{2016}]%
        {pietriga2016exploratory}
\bibfield{author}{\bibinfo{person}{Emmanuel Pietriga},
  \bibinfo{person}{Fernando Del~Campo}, \bibinfo{person}{Amanda Ibsen},
  \bibinfo{person}{Romain Primet}, \bibinfo{person}{Caroline Appert},
  \bibinfo{person}{Olivier Chapuis}, \bibinfo{person}{Maren Hempel},
  \bibinfo{person}{Roberto Mu{\~n}oz}, \bibinfo{person}{Susana Eyheramendy},
  \bibinfo{person}{Andres Jordan}, {et~al\mbox{.}}}
  \bibinfo{year}{2016}\natexlab{}.
\newblock \showarticletitle{Exploratory visualization of astronomical data on
  ultra-high-resolution wall displays}. In \bibinfo{booktitle}{\emph{Software
  and Cyberinfrastructure for Astronomy IV}}, Vol.~\bibinfo{volume}{9913}.
  International Society for Optics and Photonics, \bibinfo{pages}{99130W}.
\newblock
\urldef\tempurl%
\url{https://doi.org/10.1117/12.2231191}
\showDOI{\tempurl}


\bibitem[\protect\citeauthoryear{Prouzeau, Bezerianos, and Chapuis}{Prouzeau
  et~al\mbox{.}}{2016a}]%
        {prouzeau2016evaluating}
\bibfield{author}{\bibinfo{person}{Arnaud Prouzeau}, \bibinfo{person}{Anastasia
  Bezerianos}, {and} \bibinfo{person}{Olivier Chapuis}.}
  \bibinfo{year}{2016}\natexlab{a}.
\newblock \showarticletitle{Evaluating multi-user selection for exploring graph
  topology on wall-displays}.
\newblock \bibinfo{journal}{\emph{IEEE Transactions on Visualization and
  Computer Graphics}} \bibinfo{volume}{23}, \bibinfo{number}{8}
  (\bibinfo{year}{2016}), \bibinfo{pages}{1936--1951}.
\newblock
\urldef\tempurl%
\url{https://doi.org/10.1109/TVCG.2016.2592906}
\showDOI{\tempurl}


\bibitem[\protect\citeauthoryear{Prouzeau, Bezerianos, and Chapuis}{Prouzeau
  et~al\mbox{.}}{2016b}]%
        {prouzeau2016towards}
\bibfield{author}{\bibinfo{person}{Arnaud Prouzeau}, \bibinfo{person}{Anastasia
  Bezerianos}, {and} \bibinfo{person}{Olivier Chapuis}.}
  \bibinfo{year}{2016}\natexlab{b}.
\newblock \showarticletitle{Towards road traffic management with forecasting on
  wall displays}. In \bibinfo{booktitle}{\emph{Proceedings of the ACM
  International Conference on Interactive Surfaces and Spaces}}. ACM,
  \bibinfo{pages}{119--128}.
\newblock
\urldef\tempurl%
\url{https://doi.org/10.1145/2992154.2992158}
\showDOI{\tempurl}


\bibitem[\protect\citeauthoryear{Prouzeau, Bezerianos, and Chapuis}{Prouzeau
  et~al\mbox{.}}{2018}]%
        {prouzeau2018awareness}
\bibfield{author}{\bibinfo{person}{Arnaud Prouzeau}, \bibinfo{person}{Anastasia
  Bezerianos}, {and} \bibinfo{person}{Olivier Chapuis}.}
  \bibinfo{year}{2018}\natexlab{}.
\newblock \showarticletitle{Awareness Techniques to Aid Transitions between
  Personal and Shared Workspaces in Multi-Display Environments}. In
  \bibinfo{booktitle}{\emph{Proceedings of the ACM International Conference on
  Interactive Surfaces and Spaces}}. ACM, \bibinfo{pages}{291--304}.
\newblock
\urldef\tempurl%
\url{https://doi.org/10.1145/3279778.3279780}
\showDOI{\tempurl}


\bibitem[\protect\citeauthoryear{Reda, Johnson, Papka, and Leigh}{Reda
  et~al\mbox{.}}{2015}]%
        {reda2015effects}
\bibfield{author}{\bibinfo{person}{Khairi Reda}, \bibinfo{person}{Andrew~E
  Johnson}, \bibinfo{person}{Michael~E Papka}, {and} \bibinfo{person}{Jason
  Leigh}.} \bibinfo{year}{2015}\natexlab{}.
\newblock \showarticletitle{Effects of display size and resolution on user
  behavior and insight acquisition in visual exploration}. In
  \bibinfo{booktitle}{\emph{Proceedings of the 33rd Annual ACM Conference on
  Human Factors in Computing Systems}}. ACM, \bibinfo{pages}{2759--2768}.
\newblock
\urldef\tempurl%
\url{https://doi.org/10.1145/2702123.2702406}
\showDOI{\tempurl}


\bibitem[\protect\citeauthoryear{Renambot, Marrinan, Aurisano, Nishimoto,
  Mateevitsi, Bharadwaj, Long, Johnson, Brown, and Leigh}{Renambot
  et~al\mbox{.}}{2016}]%
        {renambot2016sage2}
\bibfield{author}{\bibinfo{person}{Luc Renambot}, \bibinfo{person}{Thomas
  Marrinan}, \bibinfo{person}{Jillian Aurisano}, \bibinfo{person}{Arthur
  Nishimoto}, \bibinfo{person}{Victor Mateevitsi}, \bibinfo{person}{Krishna
  Bharadwaj}, \bibinfo{person}{Lance Long}, \bibinfo{person}{Andy Johnson},
  \bibinfo{person}{Maxine Brown}, {and} \bibinfo{person}{Jason Leigh}.}
  \bibinfo{year}{2016}\natexlab{}.
\newblock \showarticletitle{SAGE2: A collaboration portal for scalable
  resolution displays}.
\newblock \bibinfo{journal}{\emph{Future Generation Computer Systems}}
  \bibinfo{volume}{54} (\bibinfo{year}{2016}), \bibinfo{pages}{296--305}.
\newblock


\bibitem[\protect\citeauthoryear{Robertson, Czerwinski, Baudisch, Meyers,
  Robbins, Smith, and Tan}{Robertson et~al\mbox{.}}{2005}]%
        {robertson2005large}
\bibfield{author}{\bibinfo{person}{George Robertson}, \bibinfo{person}{Mary
  Czerwinski}, \bibinfo{person}{Patrick Baudisch}, \bibinfo{person}{Brian
  Meyers}, \bibinfo{person}{Daniel Robbins}, \bibinfo{person}{Greg Smith},
  {and} \bibinfo{person}{Desney Tan}.} \bibinfo{year}{2005}\natexlab{}.
\newblock \showarticletitle{The large-display user experience}.
\newblock \bibinfo{journal}{\emph{IEEE computer graphics and applications}}
  \bibinfo{volume}{25}, \bibinfo{number}{4} (\bibinfo{year}{2005}),
  \bibinfo{pages}{44--51}.
\newblock


\bibitem[\protect\citeauthoryear{Rogers, Lim, Hazlewood, and Marshall}{Rogers
  et~al\mbox{.}}{2009}]%
        {rogers2009equal}
\bibfield{author}{\bibinfo{person}{Yvonne Rogers}, \bibinfo{person}{Youn-kyung
  Lim}, \bibinfo{person}{William~R Hazlewood}, {and} \bibinfo{person}{Paul
  Marshall}.} \bibinfo{year}{2009}\natexlab{}.
\newblock \showarticletitle{Equal opportunities: Do shareable interfaces
  promote more group participation than single user displays?}
\newblock \bibinfo{journal}{\emph{Human--Computer Interaction}}
  \bibinfo{volume}{24}, \bibinfo{number}{1-2} (\bibinfo{year}{2009}),
  \bibinfo{pages}{79--116}.
\newblock


\bibitem[\protect\citeauthoryear{Rooney, Endert, Fekete, Hornb{\ae}k, and
  North}{Rooney et~al\mbox{.}}{2013}]%
        {rooney2013powerwall}
\bibfield{author}{\bibinfo{person}{Chris Rooney}, \bibinfo{person}{Alex
  Endert}, \bibinfo{person}{Jean-Daniel Fekete}, \bibinfo{person}{Kasper
  Hornb{\ae}k}, {and} \bibinfo{person}{Chris North}.}
  \bibinfo{year}{2013}\natexlab{}.
\newblock \showarticletitle{Powerwall: int. workshop on interactive,
  ultra-high-resolution displays}. In \bibinfo{booktitle}{\emph{CHI Extended
  Abstracts on Human Factors in Computing Systems}}. ACM,
  \bibinfo{pages}{3227--3230}.
\newblock
\urldef\tempurl%
\url{https://doi.org/10.1145/2468356.2479653}
\showDOI{\tempurl}


\bibitem[\protect\citeauthoryear{Rooney and Ruddle}{Rooney and Ruddle}{2012}]%
        {rooney2012improving}
\bibfield{author}{\bibinfo{person}{Chris Rooney} {and} \bibinfo{person}{Roy
  Ruddle}.} \bibinfo{year}{2012}\natexlab{}.
\newblock \showarticletitle{Improving window manipulation and content
  interaction on high-resolution, wall-sized displays}.
\newblock \bibinfo{journal}{\emph{International Journal of Human-Computer
  Interaction}} \bibinfo{volume}{28}, \bibinfo{number}{7}
  (\bibinfo{year}{2012}), \bibinfo{pages}{423--432}.
\newblock


\bibitem[\protect\citeauthoryear{Ruddle, Thomas, Randell, Quirke, and
  Treanor}{Ruddle et~al\mbox{.}}{2016}]%
        {ruddle2016design}
\bibfield{author}{\bibinfo{person}{Roy~A Ruddle}, \bibinfo{person}{Rhys~G
  Thomas}, \bibinfo{person}{Rebecca Randell}, \bibinfo{person}{Philip Quirke},
  {and} \bibinfo{person}{Darren Treanor}.} \bibinfo{year}{2016}\natexlab{}.
\newblock \showarticletitle{The design and evaluation of interfaces for
  navigating gigapixel images in digital pathology}.
\newblock \bibinfo{journal}{\emph{ACM Transactions on Computer-Human
  Interaction (TOCHI)}} \bibinfo{volume}{23}, \bibinfo{number}{1}
  (\bibinfo{year}{2016}), \bibinfo{pages}{5}.
\newblock
\urldef\tempurl%
\url{https://doi.org/10.1145/2834117}
\showDOI{\tempurl}


\bibitem[\protect\citeauthoryear{Shuffler, Salas, and Rosen}{Shuffler
  et~al\mbox{.}}{2020}]%
        {shuffler2020evolution}
\bibfield{author}{\bibinfo{person}{Marissa~L Shuffler},
  \bibinfo{person}{Eduardo Salas}, {and} \bibinfo{person}{Michael~A Rosen}.}
  \bibinfo{year}{2020}\natexlab{}.
\newblock \showarticletitle{The evolution and maturation of teams in
  organizations: Convergent trends in the new dynamic science of teams}.
\newblock \bibinfo{journal}{\emph{Frontiers in Psychology}}
  \bibinfo{volume}{11} (\bibinfo{year}{2020}).
\newblock


\bibitem[\protect\citeauthoryear{Stasser and Abele}{Stasser and Abele}{2020}]%
        {stasser2020collective}
\bibfield{author}{\bibinfo{person}{Garold Stasser} {and}
  \bibinfo{person}{Susanne Abele}.} \bibinfo{year}{2020}\natexlab{}.
\newblock \showarticletitle{Collective choice, collaboration, and
  communication}.
\newblock \bibinfo{journal}{\emph{Annual Review of Psychology}}
  \bibinfo{volume}{71} (\bibinfo{year}{2020}), \bibinfo{pages}{589--612}.
\newblock


\bibitem[\protect\citeauthoryear{Stasser and Titus}{Stasser and Titus}{1985}]%
        {stasser1985pooling}
\bibfield{author}{\bibinfo{person}{Garold Stasser} {and}
  \bibinfo{person}{William Titus}.} \bibinfo{year}{1985}\natexlab{}.
\newblock \showarticletitle{Pooling of unshared information in group decision
  making: Biased information sampling during discussion.}
\newblock \bibinfo{journal}{\emph{Journal of personality and social
  psychology}} \bibinfo{volume}{48}, \bibinfo{number}{6}
  (\bibinfo{year}{1985}), \bibinfo{pages}{1467}.
\newblock


\bibitem[\protect\citeauthoryear{Streitz, Gei{\ss}ler, Holmer, Konomi,
  M{\"u}ller-Tomfelde, Reischl, Rexroth, Seitz, and Steinmetz}{Streitz
  et~al\mbox{.}}{1999}]%
        {streitz1999land}
\bibfield{author}{\bibinfo{person}{Norbert~A Streitz},
  \bibinfo{person}{J{\"o}rg Gei{\ss}ler}, \bibinfo{person}{Torsten Holmer},
  \bibinfo{person}{Shin'ichi Konomi}, \bibinfo{person}{Christian
  M{\"u}ller-Tomfelde}, \bibinfo{person}{Wolfgang Reischl},
  \bibinfo{person}{Petra Rexroth}, \bibinfo{person}{Peter Seitz}, {and}
  \bibinfo{person}{Ralf Steinmetz}.} \bibinfo{year}{1999}\natexlab{}.
\newblock \showarticletitle{i-LAND: an interactive landscape for creativity and
  innovation}. In \bibinfo{booktitle}{\emph{Proceedings of the SIGCHI
  conference on Human factors in computing systems}}.
  \bibinfo{pages}{120--127}.
\newblock


\bibitem[\protect\citeauthoryear{Thomas, Kannampallil, Abraham, and
  Marai}{Thomas et~al\mbox{.}}{2017}]%
        {thomas2017echo}
\bibfield{author}{\bibinfo{person}{Manu~Mathew Thomas}, \bibinfo{person}{Thomas
  Kannampallil}, \bibinfo{person}{Joanna Abraham}, {and}
  \bibinfo{person}{G~Elisabeta Marai}.} \bibinfo{year}{2017}\natexlab{}.
\newblock \showarticletitle{Echo: A large display interactive visualization of
  icu data for effective care handoffs}. In \bibinfo{booktitle}{\emph{2017 IEEE
  Workshop on Visual Analytics in Healthcare (VAHC)}}. IEEE,
  \bibinfo{pages}{47--54}.
\newblock
\urldef\tempurl%
\url{https://doi.org/10.1109/VAHC.2017.8387500}
\showDOI{\tempurl}


\bibitem[\protect\citeauthoryear{Vogt, Bradel, Andrews, North, Endert, and
  Hutchings}{Vogt et~al\mbox{.}}{2011}]%
        {vogt2011co}
\bibfield{author}{\bibinfo{person}{Katherine Vogt}, \bibinfo{person}{Lauren
  Bradel}, \bibinfo{person}{Christopher Andrews}, \bibinfo{person}{Chris
  North}, \bibinfo{person}{Alex Endert}, {and} \bibinfo{person}{Duke
  Hutchings}.} \bibinfo{year}{2011}\natexlab{}.
\newblock \showarticletitle{Co-located collaborative sensemaking on a large
  high-resolution display with multiple input devices}. In
  \bibinfo{booktitle}{\emph{IFIP Conference on Human-Computer Interaction}}.
  Springer, \bibinfo{pages}{589--604}.
\newblock


\bibitem[\protect\citeauthoryear{von Zadow, B{\"o}sel, Dam, Lehmann,
  Reipschl{\"a}ger, and Dachselt}{von Zadow et~al\mbox{.}}{2016a}]%
        {von2016miners}
\bibfield{author}{\bibinfo{person}{Ulrich von Zadow}, \bibinfo{person}{Daniel
  B{\"o}sel}, \bibinfo{person}{Duc~Dung Dam}, \bibinfo{person}{Anke Lehmann},
  \bibinfo{person}{Patrick Reipschl{\"a}ger}, {and} \bibinfo{person}{Raimund
  Dachselt}.} \bibinfo{year}{2016}\natexlab{a}.
\newblock \showarticletitle{Miners: Communication and awareness in
  collaborative gaming at an interactive display wall}. In
  \bibinfo{booktitle}{\emph{Proceedings of the ACM International Conference on
  Interactive Surfaces and Spaces}}. ACM, \bibinfo{pages}{235--240}.
\newblock
\urldef\tempurl%
\url{https://doi.org/10.1145/2992154.2992174}
\showDOI{\tempurl}


\bibitem[\protect\citeauthoryear{von Zadow, B{\"o}sel, Dam, Lehmann,
  Reipschl{\"a}ger, and Dachselt}{von Zadow et~al\mbox{.}}{2016b}]%
        {Zadow2016miners}
\bibfield{author}{\bibinfo{person}{Ulrich von Zadow}, \bibinfo{person}{Daniel
  B{\"o}sel}, \bibinfo{person}{Duc~Dung Dam}, \bibinfo{person}{Anke Lehmann},
  \bibinfo{person}{Patrick Reipschl{\"a}ger}, {and} \bibinfo{person}{Raimund
  Dachselt}.} \bibinfo{year}{2016}\natexlab{b}.
\newblock \showarticletitle{Miners: Communication and awareness in
  collaborative gaming at an interactive display wall}. In
  \bibinfo{booktitle}{\emph{Proceedings of the ACM International Conference on
  Interactive Surfaces and Spaces}}. ACM, \bibinfo{pages}{235--240}.
\newblock
\urldef\tempurl%
\url{https://doi.org/10.1145/2992154.2992174}
\showDOI{\tempurl}


\bibitem[\protect\citeauthoryear{Wallace, Iskander, and Lank}{Wallace
  et~al\mbox{.}}{2016}]%
        {wallace2016creating}
\bibfield{author}{\bibinfo{person}{James~R Wallace}, \bibinfo{person}{Nancy
  Iskander}, {and} \bibinfo{person}{Edward Lank}.}
  \bibinfo{year}{2016}\natexlab{}.
\newblock \showarticletitle{Creating your bubble: Personal space on and around
  large public displays}. In \bibinfo{booktitle}{\emph{Proceedings of the CHI
  Conference on Human Factors in Computing Systems}}. ACM,
  \bibinfo{pages}{2087--2092}.
\newblock
\urldef\tempurl%
\url{https://doi.org/10.1145/2858036.2858118}
\showDOI{\tempurl}


\bibitem[\protect\citeauthoryear{Wiese and Burke}{Wiese and Burke}{2019}]%
        {wiese2019understanding}
\bibfield{author}{\bibinfo{person}{Christopher~W Wiese} {and}
  \bibinfo{person}{C~Shawn Burke}.} \bibinfo{year}{2019}\natexlab{}.
\newblock \showarticletitle{Understanding team learning dynamics over time}.
\newblock \bibinfo{journal}{\emph{Frontiers in Psychology}}
  \bibinfo{volume}{10} (\bibinfo{year}{2019}), \bibinfo{pages}{1417}.
\newblock


\bibitem[\protect\citeauthoryear{Wigdor, Jiang, Forlines, Borkin, and
  Shen}{Wigdor et~al\mbox{.}}{2009}]%
        {wigdor2009wespace}
\bibfield{author}{\bibinfo{person}{Daniel Wigdor}, \bibinfo{person}{Hao Jiang},
  \bibinfo{person}{Clifton Forlines}, \bibinfo{person}{Michelle Borkin}, {and}
  \bibinfo{person}{Chia Shen}.} \bibinfo{year}{2009}\natexlab{}.
\newblock \showarticletitle{WeSpace: the design development and deployment of a
  walk-up and share multi-surface visual collaboration system}. In
  \bibinfo{booktitle}{\emph{Proceedings of the SIGCHI Conference on Human
  Factors in Computing Systems}}. \bibinfo{pages}{1237--1246}.
\newblock


\end{thebibliography}

\end{document}